\documentclass[11pt]{article}
\usepackage{arxiv}

\usepackage{graphicx}
\usepackage{caption}
\usepackage{subcaption}
\usepackage{wrapfig}
\usepackage{svg}

\usepackage{amsmath}
\usepackage{amssymb}
\usepackage{amsthm}
\usepackage{mathrsfs} 
\usepackage{bm}  
\usepackage{centernot}  

\usepackage[linesnumbered, ruled, vlined]{algorithm2e}
\usepackage{color}
\usepackage{mathtools}  

\usepackage[backend=biber,sorting=nyt,firstinits=true,maxnames=10000]{biblatex}
\usepackage{enumitem}
\usepackage{etex,etoolbox}
\usepackage{hyperref}
\usepackage{todonotes}
\bibliography{../../refs}
\usepackage{booktabs}
\usepackage{thmtools}
\usepackage{environ}
\theoremstyle{plain}  

\usepackage[conf={restate}]{proof-at-the-end}

\newtheorem{theorem}{Theorem}[section]

\theoremstyle{definition}
\newtheorem{remark}{Remark}[section]
\newtheorem{definition}{Definition}[section]

\newtheorem{assumption}{Assumption}[section]

\def\defeq{\overset{\Delta}{=}}  
\def\intr{\mathsf{int}}  
\def\ran{\mathsf{ran}}  

\def\d{\mathsf{d}}  

\def\E{\mathbb{E}}  
\def\P{\mathbb{P}}  

\def\R{\mathbb{R}}  

\def\ind{\mathbf{1}}  
\def\<{\langle}  
\def\>{\rangle}  

\def\B{\mathcal{B}}  
\def\A{\mathcal{A}}  
\def\Tt{\mathcal{T}_t}  
\def\Tk{\mathcal{T}_{T_k}}  
\def\L{\mathcal{L}}  

\newcommand{\RNum}[1]{\uppercase\expandafter{\romannumeral #1\relax}}  

\def\Wb{\mathbf{W}}

\def\Lb{\overline{\Lambda}}

\def\Tb{\widetilde{T}}
\def\Wt{\widetilde{W}}
\def\Lt{\widetilde{\Lambda}}
\def\dt{\d t}

\def\G{\mathcal{G}}  

\newcommand{\PK}[1]{P_#1}  

\graphicspath{{figures/}}

\title{Real-time Bidding for Time Constrained Impression Contracts in
  First and Second Price Auctions -- Theory and Algorithms}

\author{R. J. Kinnear\\
  \small\href{mailto:ryan@kinnear.ca}{ryan@kinnear.ca} \and
  R. R. Mazumdar\\
  \small\href{mailto:mazum@uwaterloo.ca}{mazum@uwaterloo.ca} \and
  P. Marbach\\
  \small\href{mailto:marbach@cs.utoronto.edu}{marbach@cs.utoronto.edu}}

\begin{document}
\maketitle
\begin{abstract}
  We study the optimal behavior of a bidder in a real-time auction
subject to the requirement that a specified collections of
heterogeneous items be acquired within given time constraints.  The
problem facing this bidder is cast as a continuous time optimization
problem which we show can, under certain weak assumptions, be reduced
to a \textit{convex} optimization problem.  Focusing on the standard
first and second price auction mechanisms, we first show, using convex
duality, that the optimal (infinite dimensional) bidding policy can be
represented by a single \textit{finite} vector of so-called
``pseudo-bids''.  Using this result we are able to show that, in
contrast to the first price auction, the optimal solution in the
second price case turns out to be a very simple piecewise constant
function of time.  Moreover, despite the fact that the optimal
solution for the first price auction is genuinely dynamic, we show
that there remains a close connection between the two cases and that,
empirically, there is almost no difference between optimal behavior
in either setting.  Finally, we detail methods for implementing our
bidding policies in practice with further numerical simulations
illustrating the performance.

\end{abstract}


\paragraph{Acknowledgement}
We acknowledge the support of the Natural Sciences and Engineering
Research Council of Canada (NSERC), [funding reference number
518418-2018].  Cette recherche a été financée par le Conseil de
recherches en sciences naturelles et en génie du Canada (CRSNG),
[numéro de référence 518418-2018].


\section{Introduction}
Since at least 2016 online advertising, which includes ads in
cellphone apps, website banner ads, search engine keyword ads,
etc. has been the dominant advertising segment \cite{iab2018iab}. One
of the drivers of this dominance is not only the fact that so many
people are now online, but that the technology enables more accurate
targeting.  That is, advertisers can accurately target their messaging
to the groups or individuals to which the message is most relevant.
This technology enables advertisers to raise awareness of their
products, website or app operators (i.e., ``publishers'') to monetize
their audiences, and internet users to access a wealth of gratis
content.

There are many different mechanisms through which publishers can sell
ad space.  For example, sponsored search is a well known mechanism for
search engines where certain keywords (e.g. ``mortgage'', ``used
car'', etc.) trigger the display of relevant sponsored results.  More
traditional fixed contracts (e.g., for the display of fixed banner
ads) between advertisers and publishers also remain common,
particularly for businesses which need to be careful about their image
and want to maintain complete control over what their brand is
associated with.  Finally, a significant portion of total advertising
revenue is generated through a mechanism called real-time bidding
(RTB), which is the mechanism at the focus of this paper.

RTB is facilitated by auction exchanges (e.g., Google AdX
\cite{mansour2012doubleclick}) where the arrival of each user to a
publisher website or app triggers a sealed-bid auction (the bid is
hidden from other participants) wherein advertisers bid on the right
to display their ad to the particular user.  Generically, we will
think of this as a single ``item'', and the known characteristics
(e.g., age, interests, etc.) of the visitor as the item's ``type''.
Advertisers decide upon which types they wish to target, and bid on
these items when they arrive at the auction exchange.

In this paper we study a problem of interest to intermediaries or
brokers known as Demand Side Platforms (DSPs), see
e.g. \cite{wang2017display}. Advertisers contract with these DSPs to
bid on the RTB market on their behalf. Specifically, we consider DSPs
that have entered into contracts which require a specified number of
items of particular types be won within a given time deadline.  We do
not allow for the contract requirements to be subject to post hoc
negotiation -- once the contract is accepted the DSP is obligated to
meet the requirements.  This benefits the advertiser by allowing them
to offload the risk associated with price changes in the market, and
gives the DSPs the opportunity to profit from managing this risk on
behalf of multiple advertisers.  Thus, the objective of the DSP is to
win the required items in RTB at the lowest possible cost (since the
difference between the contract value and its cost of acquiring the
ads is the profit).  This is a natural setting to consider as there is
a reasonable expectation of efficiency gains when the demands of a
large number of advertisers are aggregated by intermediaries.

The dominant auction mechanisms in RTB are \textit{first price} and
\textit{second price} auctions.  In either case, the item is awarded
to the highest bidder, but the difference lies in what they pay: in a
first price auction, the winning bidder pays their actual bid, whereas
in a second price auction, the winning bidder pays whatever the
\textit{second highest} bid was.  In both auctions, the bids are
sealed (other participants don't observe them, even after an item is
allocated) so the only source of information about prices are from
available historical data, and the knowledge gained from the bidder's
direct participation.  Standard overviews of auction theory are
provided by \cite{krishna2009auction, menezes2005introduction}.

\paragraph{Related Work}
Real-time bidding is commonly studied from at least two main
perspectives: the descriptive analysis of the market itself (e.g.,
\cite{iyer2014mean, balseiro2015repeated, balseiro2019learning}),
where questions of game theoretic equilibria are at the forefront; and
the consideration of normative problems of optimal bidding.  Our
perspective is normative.  We solve an optimal bidding and allocation
problem by providing algorithms which produce a plan over a given time
horizon specifying the optimal average rate that items should be
acquired, and to which contracts (or campaigns) they should be
allocated.  Similar allocation problems have been considered from the
publisher's viewpoint \cite{chen2011real, radovanovic2012risk} and the
advertiser's \cite{zhang2015statistical, marbach_bidding_2020}.
However, aside from \cite{marbach_bidding_2020}, no other papers, to
our knowledge, have considered contract management aspects with hard
constraints from the DSP perspective.  Many works have studied the
problem of short time horizon adaptation, for instance the works of
\cite{ghosh2009adaptive, zhang2016feedback, karlsson2014adaptive,
  karlsson2017plant} have applied classical control theory methods
(e.g., PID controllers) and those of \cite{gummadi2011optimal,
  cai2017real, wu2018budget} MDP formulations and reinforcement
learning.  The goal of these works are to maintain or maximize certain
performance indicators, e.g., click through rates or purchasing
decisions.  As such, these works can be seen as considering a faster
time scale in comparison to our problem.  Optimal bidding problems
over a long time horizon are considered by \cite{fernandez2017optimal,
  grislain2019recurrent} via application of stochastic optimal
control, where the goal is to maximize the total valuation of items
attained over the period, subject to budget constraints.  An important
consideration has thus also been the estimation of item valuations,
and how to map valuations into bids, see e.g., \cite{perlich2012bid,
  zhang2014optimal}.  These are distinct from the present work since
we do not appeal to any notion of item valuation, indeed, a valuation
can only be derived from the contract requirements.  Our formulation
is also dual to many of these as we seek to minimize the cost of
obtaining a given number of items, as opposed to maximizing the
utility of items obtained within a given budget.  A similar class of
problems to our optimal acquisition problem is that of revenue
management (see e.g., \cite{wang2020constant, bumpensanti2020re})
wherein a merchant needs to set prices for their wares in order to
optimally liquidate their inventory.  Aside from the superficial
distinction of sell-side versus buy-side, the contract management
problem studied in the present paper involves substantially different
constraints.  Finally, many papers have also studied separate, but
important, statistical estimation problems (e.g. \cite{cui2011bid,
  wu2015predicting, zhang2016bid, zhu2017gamma, wu2018deep,
  ghosh2020scalable}) concerned with estimating the prevailing prices
of arriving items.

\paragraph{Contributions}
This paper studies an optimal bidding and allocation problem similar
to the earlier work of \cite{marbach_bidding_2020} for second price
static auctions, but with the additional consideration of temporal
dynamics and differing contract deadlines.  The main contributions of
this paper over those of \cite{marbach_bidding_2020} are threefold:
Firstly, the recognition that the problem is in fact naturally
\textit{convex}, this leads to an analysis which is general enough to
unify both the first and second price auctions, as well as to result
in an insightful duality theory.  Secondly, a clear development of
practical and simple algorithms for implementing solutions to the
optimal bidding and allocation problem are provided, along with
detailed simulation evidence and practical methods for adaptation to a
stochastic environment.  And finally, we show that there is a close
connection between the first and second price cases, both
theoretically as well as through experimental evidence.

\paragraph{Outline}
The outline of the paper is as follows: We formally introduce the
problem in Section \ref{sec:optimal_bidding}. In Section
\ref{sec:convex_acquisition_costs}, we show that it can be
reformulated as an infinite dimensional convex optimization problem
(albeit with some additional assumptions in the first price case, see
Proposition \ref{prop:convex_acquisition_costs}).  The convex
formulation and a strong duality theorem is provided in Section
\ref{sec:optimal_bidding_in_ct}, where we also show that the optimal
time dependent bids are fully determined by a finite set of parameters
deemed \textit{pseudo bids}.  Using duality, we show in Section
\ref{sec:connections} that there is a close connection between the
first and second price case; this connection is expanded upon in the
simulations of Section \ref{sec:simulation}, where experimental
evidence demonstrates that there is little practical difference
between the solutions calculated in the two cases.  In terms of
implementation, the dual problem in Section
\ref{sec:optimal_bidding_in_ct} is finite and can be solved to attain
exact solutions, but doing so requires computationally expensive
integration steps; this motivates a simpler polyhedral approximation
method in Section \ref{sec:approximation}, where we also bound the
error between the approximation and the true optimal solution.  Using
the methods of Section \ref{sec:approximation}, an empirical study is
carried out in Section \ref{sec:simulation} using real RTB data.
Motivated by the results of this simulation, Section
\ref{sec:improving_fulfillment} examines additional practical methods
for online bid updates and for accounting for uncertainty in any
estimated quantities.  Finally, Section \ref{sec:conclusion}
concludes.  Proofs are provided in the appendix.

\section{The General Optimal Bidding Problem}
\label{sec:optimal_bidding}
We consider the problem of bidding in a sequential auction with $M$
available item types $j \in [M] \defeq \{1, 2, \ldots, M\}$ and where
we are required to fulfill $N$ contracts by winning a specified number
of items of given types.  Specifically, the contract $i \in [N]$ is a
$3$-tuple $(\A_i, C_i, T_i)$, where $\A_i \subseteq [M]$ is the set of
item types that can be allocated towards the contract, $C_i$ is the
number of items (of any type in $\A_i$) that are needed, and $T_i$ is
the time deadline by which the items must be obtained.  Another
collection of sets, $\B_i$, is uniquely determined by the condition
$i \in \A_j \iff j \in \B_i$, and the sets $\A_i, \B_j$ therefore
determine a bipartite graph $\G$ on $[N] \times [M]$ encoding which
item types can be allocated to which contracts, and conversely, which
contracts can be satisfied by which item types.  It will be convenient
to write $(i, j) \in \G$ to mean that $i, j$ is an edge in this graph.

We will suppose without loss of generality that
$0 < T_1 \le T_2 \le \cdots \le T_N$.  Contracts stipulating more
complicated item requirements (e.g., at least $100$ of type $j$, and
an additional $50$ of either $j$ or $j'$) can be implemented by
introducing multiple contracts with the same deadlines.  Denoting
$T \defeq T_N$ for the end of the problem horizon and $T_0 \defeq 0$
will also be notationally convenient.  The time $T_i$ indicates when
contract $i$ ``exits'' the problem, and similarly it will be
convenient to write $T^j = \underset{i \in \B_j}{\text{max }} T_i$ for
the time that items of type $j$ exit, i.e., are no longer needed.

\begin{remark}[Non-zero Initial Times]
  The framework can also incorporate introducing new contracts part
  way through the fulfillment of existing ones by ``resetting'' the
  existing contracts by subtracting the current time and already
  obtained supply from their requirements, and then adding the new
  contracts to the collection before recalculating the optimal bids.
  This is further explored in Section \ref{sec:receding_horizon} for
  the purpose of adapting to a stochastic environment.  Managing
  contracts with starting times other than $0$ is also not a major
  difficulty: each contract must now come along with a starting time
  $S_i$ \textit{and} it's deadline $T_i$, the ultimate result being
  that integrals over the interval $[0, T_i]$ would need to be taken
  over $[S_i, T_i]$, and the dual problem considered in Section
  \ref{sec:optimal_bidding_in_ct} will have additional variables.  In
  some applications (e.g., processor speed scaling
  \cite{gaujal2005shortest}), there are algorithms which can take
  advantage of special structure in the arrivals of jobs (e.g., that
  the jobs be first-in first-out: $S_i \le S_j \implies T_i \le T_j$)
  to achieve performance improvements, but such structure does not
  impact the methods we develop here.
\end{remark}

\begin{remark}[Item types]
  That a contract may be fulfilled with any arrangement of items in
  $\A_i$ (e.g., $C_i$ items of a single type $j \in \A_i$ is just as
  good as an even distribution over all items in $\A_i$) appears at
  first as an unusual modelling artifact.  However, it need not be the case
  that the counter party can distinguish between item types.  For
  example, consider two contracts: contract $1$ targets anyone aged
  $30-40$, and contract $2$ targets anyone aged $20-40$.  This would
  result in the DSP recognizing two types of items: people aged
  $20-30$ and people aged $30-40$.  Contract $2$ can indeed be
  fulfilled just as well with either type, but contract $1$ can be
  fulfilled only with the second type.  
\end{remark}

\subsection{Supply Rate Curves and Cost Functions}
The model of how items are acquired relies on a function $W_j(x, t)$,
called a \textit{supply curve}, associated to each item type $j$ and
quantifying, instantaneously at time $t$, the expected number of items
that would be won by bidding $x$.  This function derives from the
assumption that our bidder is a \textit{price taker}, i.e., has no
substantial market impact.

The function $W_j(x, t)$ should be thought of us an un-normalized
(cumulative) distribution function with respect to the bid argument
$x$, but as a rate with respect to the time argument $t$.  This
function needs to be estimated from available historical data (see
e.g., ~\cite{cui2011bid, wu2015predicting, zhang2016bid,
  wang2016functional, ghosh2020scalable} for work on this problem),
and represents what is essentially a fluid approximation of the
``landscape'' of competing bids.  It is important to emphasize that
these curves represent \textit{averages}, but that the environment is
stochastic in actuality.  Assuming that we only have access to the
first order average statistics leads to a simpler problem, while at
the same time also being a \textit{weaker} assumption than if we were
to develop a full stochastic model that assumes further information
about higher order statistics.  Nevertheless, we will see in Section
\ref{sec:improving_fulfillment} some practical methods for adapting to
the stochastic environment.

We remark that the purpose of including time-dependence is to enable
the consideration of forecast changes in supply availability; in
particular, there are natural cycles present in the average available
supply throughout time: see e.g., \cite{yuan2013real}, as well as
Figure \ref{fig:supply_curve_estimates}.

Additionally, we must introduce a \textit{cost function} $f_j(x, t)$
which quantifies, instantaneously at time $t$, the expected cost of
bidding $x$ on every arriving item of type $j$.  This function
naturally depends on the auction mechanism, but also on the supply
rate curve $W_j$, since the total amount spent depends upon the number
of items which are actually won.  In order to lighten the notation, we
will often suppress the $j$ index and $t$ argument, writing
$W(x), f(x)$ etc. when it is clear from, or irrelevant to, the
context.

For a first price auction, the cost function is the product of the bid
and the supply rate curve, since you pay your bid:

\begin{equation}
  \label{eqn:1pa_cost}
  f^{1st}(x) = xW(x)\ind_{\R_+}(x),
\end{equation}

where

\begin{equation*}
  \ind_A(x) = \left\{
    \begin{array}{lr}
      1 & x \in A\\
      0 & \text{otherwise}
    \end{array}\right.,
\end{equation*}

is the indicator for the set $A$.

In the second price case, the supply rate curve becomes an integrator,
since we only pay the highest competing bid:

\begin{equation}
  \label{eqn:2pa_cost}
  \begin{aligned}
    f^{2nd}(x)
    &= \int_{0}^xu\d W(u)\ind_{\R_+}(x)\\
    &= \bigl[xW(x) - \int_{0}^x W(u)\d u\bigr]\ind_{\R_+}(x),
  \end{aligned}
\end{equation}

where the first integral is to be understood as a Lebesgue-Stieltjes
integral (see e.g. \cite[ch. 6]{stein2009real}) and we have integrated
by parts.  The inequality

\begin{equation}
  f^{2nd} \le f^{1st},
\end{equation}

is evident from the above since
$f^{2nd}(x) = f^{1st}(x) - \bigl[\int_{0}^x W(u)\d u\bigr]\ind_{\R_+}(x)$.

We will consider both first price and second price auctions as the two
cases are amenable to a unified treatment.  However, we often place
our focus on the first price case, as its analysis requires an
additional (albeit rather weak) assumption.  We will occasionally omit
the superscript $1st$ or $2nd$ (sometimes written as a subscript)
depending on whether or not we want to emphasize a particular case.
Moreover, it is important to recognize that the DSP must estimate the
curve $W(x)$ in the auction they will actually participate in.  Due to
strategic differences between the first and second price cases, these
functions will not be the same across auction types, but our analysis
has no need to distinguish between them.

We make the following basic assumptions.

\begin{assumption}[Supply Rate Curve Properties]
  \label{ass:W_f_properties}
  For every $t \in \R_+$, the functions $x \mapsto W_j(x, t)$ are
  strictly positive and continuous on $\R$, as well as being either
  strictly monotone increasing on an interval
  $(-\infty, \bar{x}_j] \subseteq \R$ and flat thereafter, or strictly
  monotone increasing on $\R$ and unbounded.  Moreover,
  $\bar{s}_j(t) \defeq \text{sup}_xW_j(x, t)$ is either $+\infty$ or
  attained at some $\bar{x}_j(t) \in \R$.  Finally,
  $W_j(x, t) \rightarrow 0$ as $x \rightarrow -\infty$.
\end{assumption}

The reasons for these assumptions, and whether or not they can be
relaxed, is discussed in the following sequence of remarks.

\begin{remark}[Attainment]
  That the supremum of $W$ must be obtained when finite is not a
  benign assumption.  It technically excludes many models of the form
  $W(x, t) = \lambda(t)W(x)$ where $\lambda(t)$ is a supply rate and
  $W(x)$ is a cumulative distribution function with unbounded support
  (e.g., $1 - e^{-\gamma x}$).  Of course, this is essentially only a
  theoretical restriction, and such models are reasonable in practice
  since one could choose some large but fixed maximum allowable bid
  $\bar{x}$ and truncate.  If the supremum is not attained when
  finite, then the inverse function $W^{-1}$ is not lower
  semi-continuous, which is an important property for rigorous proofs.
\end{remark}

\begin{remark}[Randomized Bidding]
  It is known that there are benefits to \textit{randomized bidding}
  \cite{karlsson2016control}.  That is, if $\sigma > 0$ is a
  parameter, and $x(t)$ is the estimated optimal bid at time $t$, then
  the \textit{actual} bid submit to the auction house should be the
  random quantity $b \sim \mathcal{N}(x(t), \sigma^2)$.  In this case,
  \textit{nominal} bids below $0$ still have some probability of
  inducing an \textit{actual} bid above $0$.  The $\sigma$ parameter
  can, for example, be chosen to correspond to the bandwidth of kernel
  density estimates of supply rate curves, a natural estimation
  procedure for this application.  Thus, a nominal bid of
  $x \rightarrow -\infty$ can be understood as ensuring that the
  probability of the actual bid exceeding the floor price is
  converging to $0$.
\end{remark}

\begin{remark}[Continuity]
  Following the discussion of the previous remark, randomized bidding
  also leads naturally to supply rate curves which are \textit{smooth}
  (possessing higher order derivatives), for exactly the same reason
  that Gaussian kernel density estimates are smooth.  Indeed, it would
  be reasonable to assume that supply rate curves are smooth, but we
  only \textit{require} simple continuity.  In the work of
  \cite{marbach_bidding_2020}, supply curves are initially assumed
  merely to be only right-continuous; however, it is then shown that
  optimal bidders \textit{must} randomize, which makes the curves
  effectively continuous.
\end{remark}

\subsubsection{Estimation of Supply Rate Curves}
In order to provide further context, we briefly discuss estimates of
supply curves obtained from the IPinYou dataset \cite{zhang2014real,
  liao2014ipinyou}.  We are not focused on the statistical estimation
problem, and we therefore use simple methods.  Further detail is
provided in the appendix.

\begin{figure}
  \centering
 
  \begin{subfigure}[b]{0.4\textwidth}
    \includegraphics[width=\textwidth]{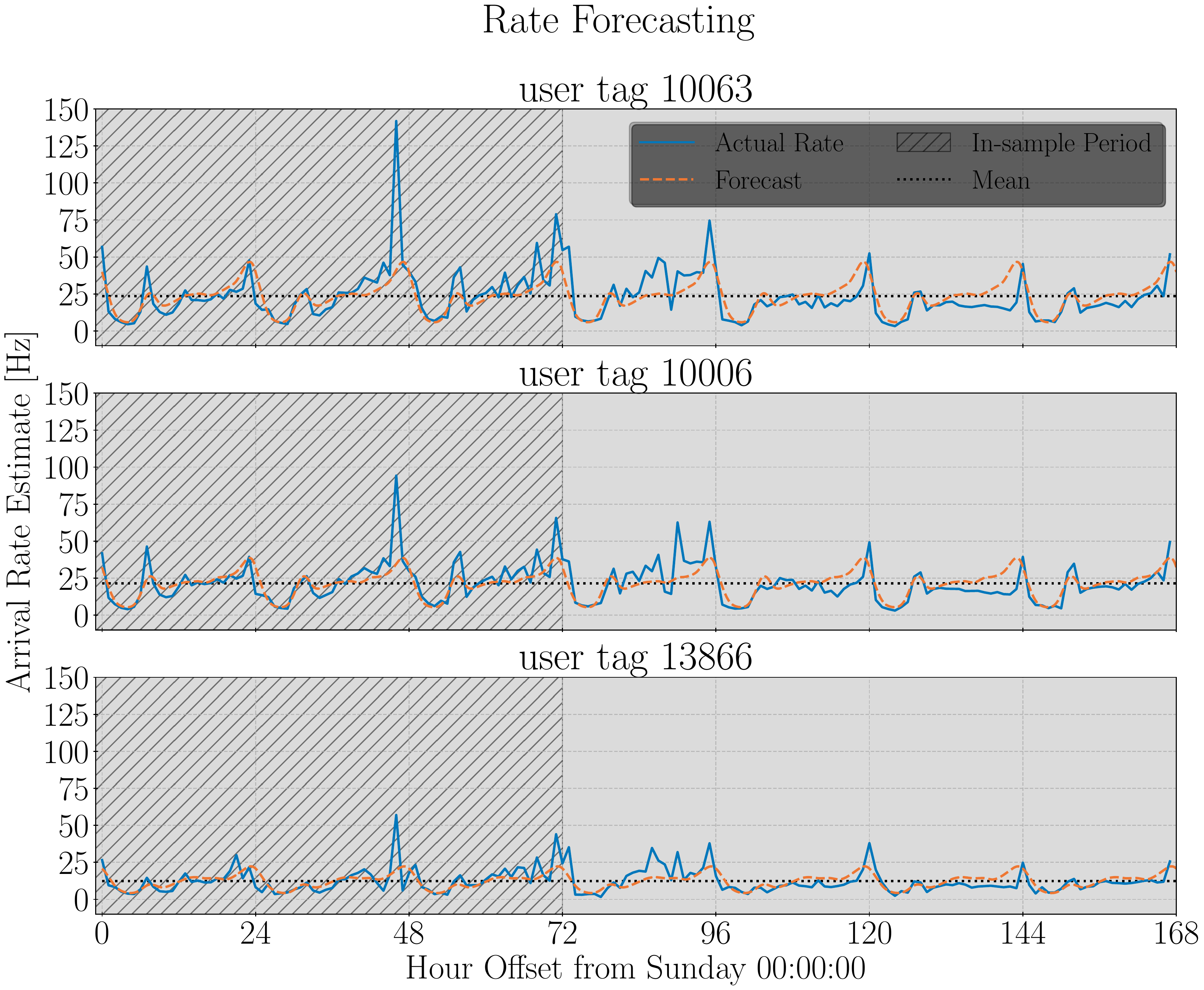}
    \caption{Supply Rate $\lambda(t)$}
    \label{fig:supply_rate_estimate}
  \end{subfigure}
  \begin{subfigure}[b]{0.4\textwidth}
    \includegraphics[width=\textwidth]{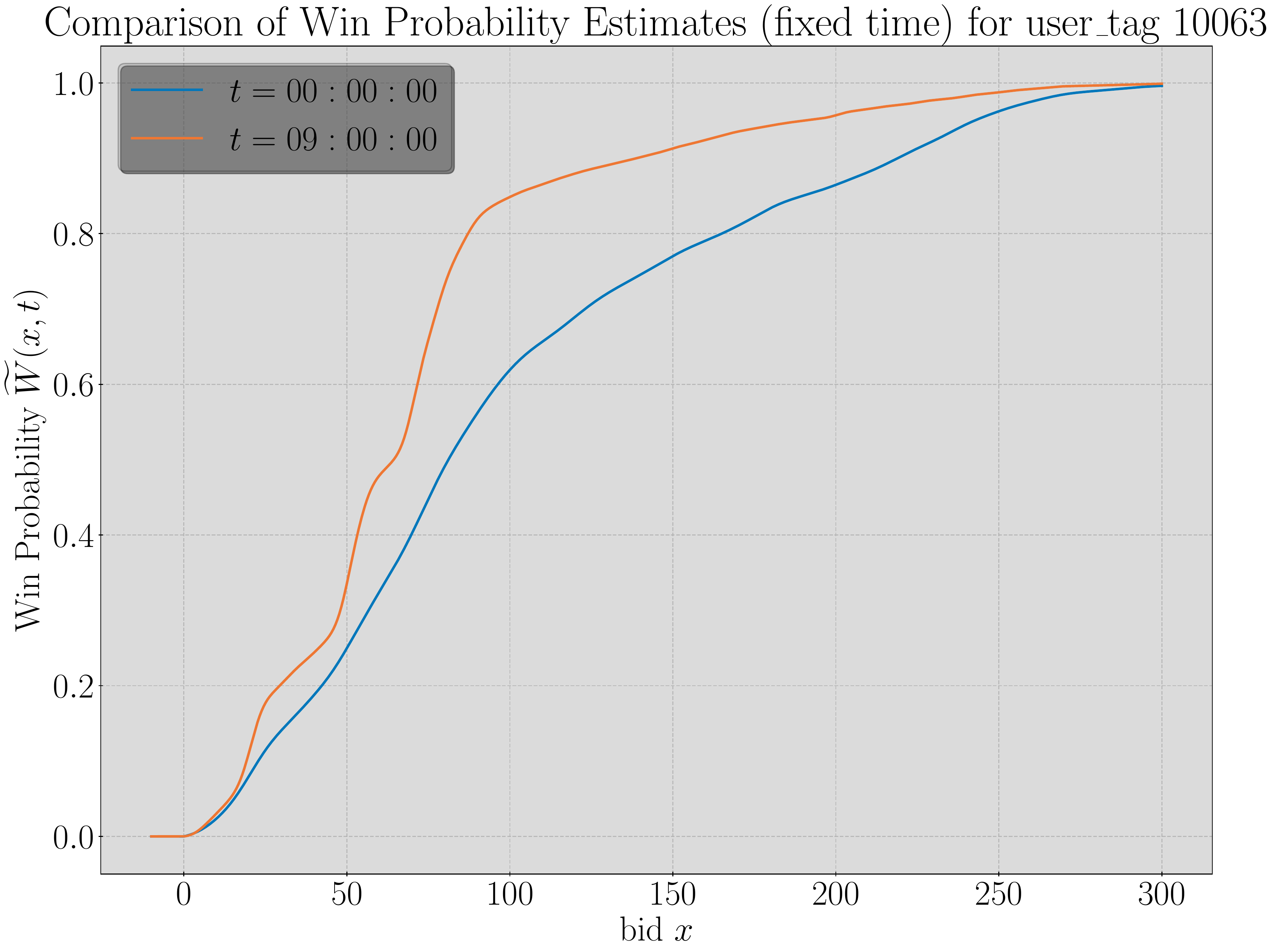}
    \caption{Win Probability Estimates}
    \label{fig:win_prob_estimate}
  \end{subfigure}
  \caption{Estimated Supply Rate Curves and Costs}
  \label{fig:supply_curve_estimates}


  {Supply rate and win probability curves estimated from iPinYou data.
    $(a)$ Item arrival rates and the corresponding forecasts.  The
    hatched region indicates an in-sample period with the remainder
    being out-of-sample.  $(b)$ Estimates of the win probability
    function $\widetilde{W}(x, t)$ are obtained via Gaussian Kernel
    Density Estimates.}
\end{figure}

In this example, the supply curve $W(x, t)$ is estimated by combining
estimates of the supply rate (the rate of items arriving), denoted by
$\lambda(t)$, and the probability of winning an item at a certain
price.  We denote the probability of winning by $\Wt(x, t)$, and the
supply rate curve estimate is then taken to be
$W(x, t) = \lambda(t) \Wt(x, t)$.  Figure
\ref{fig:supply_rate_estimate} depicts representative examples of rate
estimates and clearly depicts the typical daily cycles in supply
rates.  Figure \ref{fig:win_prob_estimate} depicts examples of
$\Wt(x, t)$ estimates.  The curves are positive, continuous, monotone
increasing, and truncated to a bid of $x = 300$, c.f., Assumption
\ref{ass:W_f_properties}.

\subsection{Problem Definition}
Our goal is to find a \textit{bid path} $x(t) \in \R^{N \times M}$ and
\textit{allocation path} $\gamma(t) \in [0, 1]^{N \times M}$ which
solves the following problem:

\begin{equation}
  \label{eqn:ct_problem}
  \begin{aligned}
    \underset{x, \gamma}{\text{minimize}} \quad & \sum_{i = 1}^N\int_{0}^{T_i} \Bigl[\sum_{j \in \A_i}\gamma_{ij}(t)f_j(x_{ij}(t), t)\Bigr]\d t\\
    \textrm{subject to} \quad & \sum_{j \in \A_i}\int_0^{T_i}\gamma_{ij}(t)W_j(x_{ij}(t), t) \d t\ge C_i\\
    \quad & \sum_{i \in \B_j} \gamma_{ij}(t) \le 1, \gamma_{ij}(t) \ge 0,
  \end{aligned}
  \tag{$TP$}
\end{equation}

where it is to be understood that constraints containing unqualified
indices are to hold over their whole range.  That is,
$\gamma_{ij}(t) \ge 0$ is short for:
$\forall (i, j) \in \G\ \forall t \in [0, T_i]\ \gamma_{ij}(t) \ge 0$.
Additionally, functions $\gamma_{ij}(t), x_{ij}(t)$ etc. are treated
as $0$ or undefined outside of $t \in [0, T_i]$.  The notation

\begin{equation}
  \Tt \defeq \{i \in [N]\ |\ t < T_i\}\\
\end{equation}

indicating the contracts active up to time $t$ will be useful in
keeping track of active contracts.

The variable $\gamma_{ij}(t)$, referred to as the \textit{allocation},
parameterizes the proportion of items of type $j$ which will be bid on
on behalf of contract $i$.  Similarly, $x_{ij}(t)$ indicates the bid
that should be submit for an item of type $j$ on behalf of contract
$i$.  Concretely, upon the arrival of an item of type $j$ at time $t$,
the DSP bids on the item with probability
$\Gamma_j(t) = \sum_{i \in \B_j} \gamma_{ij}(t)$, and if they have
chosen to bid they do so on contract $i$ with probability
$\gamma_{ij}(t) / \Gamma_j(t)$, and finally place the bid $x_{ij}(t)$
on the item.  We have here allowed for the possibility of placing
different bids depending upon which contract the item will be
allocated towards, however, it will be shown in the sequel that this
is never necessary and that Problem \eqref{eqn:ct_problem} could have
been written with bids $x_j(t)$ indexed only by $j$ and $t$, rather
than the triple $i, j, t$.

We set the stage for our solution of Problem \eqref{eqn:ct_problem}
with some final conventions.

\begin{definition}[Conventions]
  We write $W_j^{-1}(x, t)$ for the inverse function of
  $x \mapsto W_j(x, t)$.  The inverse is guaranteed to exist on the
  range of $W_j$ since the function is strictly monotone.  Outside the
  range, the inverse is understood to be $+\infty$, i.e.,
  $W_j^{-1}(s, t) = \infty$ if there is no $x \in \R$ such that
  $W_j(x, t) = s$.  Typically, we will use $x$ for bids and $s$ for
  units of supply.
\end{definition}

\subsection{Convex Acquisition Costs}
\label{sec:convex_acquisition_costs}
The key to understanding Problem \eqref{eqn:ct_problem} is to consider
\textit{acquisition cost} functions:

\begin{equation}
  \Lambda_j(s, t) \defeq f_j \circ W_j^{-1}(s, t)
\end{equation}

which quantify, instantaneously at time $t$, the average cost of
acquiring supply at the rate $s$.

In the second price case, the acquisition cost function is
\textit{always} convex.  Indeed, this can be seen by a change of
variables $y = W(u)$

\begin{align*}
  \Lambda^{2nd}(s) 
  &= \ind_{\R_+}(W^{-1}(s))\int_0^{W^{-1}(s)}u\d W(u)\\
  &= \ind_{\{s \ge W(0)\}}\int_{W(0)}^s W^{-1}(y)\d y,
\end{align*}

which is convex since its (one-sided) derivative on the domain is
monotone non-decreasing.  The interpretation of this result is that
the marginal acquisition costs are increasing
\cite{leblanc1974transportation}.  This is entirely natural for second
price auctions since you pay the same amount for items that would be
won by bids of $x$, whether or not you're bidding exactly this amount,
or something greater.

For the case of first price auctions, while total costs are
necessarily always greater for greater bids, the \textit{marginal}
costs need not necessarily be increasing and therefore convexity
does not always hold.  In order to understand when it does, we need to
discuss the idea of log-concavity.  Recall that a function $W$ is
log-concave if $\ln \circ W$ is concave and that any concave function
is log-concave, but that the converse is false.  The following notion
of concavity, which includes both concavity and log-concavity, is
utilized in our analysis.

\begin{definition}[$\alpha$-concavity]
  \label{def:quasi_log_concave}
  Define, for $\alpha \ge 0$, $x > 0$ the function

  \begin{equation*}
    \ell_\alpha(x) \defeq \int_1^x \frac{1}{t^\alpha}\d t =
    \left\{\begin{array}{lr}
             \ln x & \alpha = 1\\
             \frac{x^{1 - \alpha} - 1}{1 - \alpha} & \text{otherwise}
           \end{array}\right.,
  \end{equation*}

  where in particular $\ell_2(x) = 1 - 1/x$.  We will say that a
  positive function $W: \R \rightarrow (0, \infty)$ is (strictly)
  $\alpha$-concave if $\ell_\alpha \circ W$ is (strictly)
  concave.  In particular, $W$ is log-concave if $\alpha = 1$ and
  concave if $\alpha = 0$.
\end{definition}

The following observation tells us that for any $\alpha > 1$,
$\alpha$-concavity is a weaker requirement than log-concavity, which
is in turn weaker than concavity.

\begin{theoremEnd}[restate, end, category=aux, no link to proof]{proposition}[Hierarchy of $\alpha$-concavity]
  \label{prop:alpha_log_concavity}
  For $0 \le \alpha < \beta$, if $W$ is $\alpha$-concave, then it is
  also $\beta$-concave.
\end{theoremEnd}
\begin{proofEnd}
  First we check that $\ell_\beta \circ \ell_\alpha^{-1}(x)$ is both
  monotone increasing and concave.  This follows if the first
  derivative

  \begin{equation*}
    \frac{\d}{\d x}\ell_\beta \circ \ell_\alpha^{-1}(x)
    = \frac{\ell_\beta' \circ \ell_\alpha^{-1}(x)}{\ell_\beta' \circ \ell_\alpha^{-1}(x)},
  \end{equation*}

  is positive and monotone non-increasing; which, since
  $\ell_\alpha^{-1}$ is itself monotone non-decreasing, follows if

  \begin{equation*}
    \frac{\ell_\beta'(x)}{\ell_\alpha'(x)}
  \end{equation*}

  is both positive and monotone decreasing.  Since
  $\ell_\alpha'(x) = x^{-\alpha}$ this function is
  $x^{\alpha - \beta}$, which is positive on the domain $x > 0$ and
  decreasing if $\alpha < \beta$.

  Now, we check concavity of $\ell_\beta \circ W$ directly from the
  definition, for $t \in (0, 1)$:

  \begin{align*}
    \ell_\beta \circ W(tx + (1 - t) y)
    &= \ell_\beta \circ \ell_\alpha^{-1} \circ \ell_\alpha \circ W(tx + (1 - t)y)\\
    &\overset{(a)}{\ge} \ell_\beta \circ \ell_\alpha^{-1}\bigl(t\ell_\alpha \circ W(x) + (1 - t)\ell_\alpha \circ W(y) \bigr)\\
    &\overset{(b)}{\ge} t\ell_\beta \circ \ell_\alpha \circ \ell_\alpha^{-1} \circ W(x) + (1 - t)\ell_\beta \circ \ell_\alpha^{-1} \circ \ell_\alpha \circ W(y)\\
    &= t \ell_\beta \circ W(x) + (1 - t) \ell_\beta \circ W(y),
  \end{align*}

  where $(a)$ follows by the assumed concavity of
  $\ell_\alpha \circ W$ and the monotonicity of
  $\ell_\beta \circ \ell_\alpha^{-1}$ while $(b)$ from the concavity
  of $\ell_\beta \circ \ell_\alpha^{-1}$.
\end{proofEnd}







We can now characterize the conditions under which the cost of
acquisition function $\Lambda_j(s, t)$ is convex in $s$ for first
price auctions.  As discussed above, a similar statement holds for the
second price case, and doesn't require the log-concavity assumption.

\begin{theoremEnd}[end, restate, category=main, no link to proof]{proposition}[Convex Acquisition Costs -- First Price Case]
  \label{prop:convex_acquisition_costs}
  Suppose that a supply rate curve $W(x)$ is strictly $2$-concave.  Then in
  a first price auction, the extended acquisition function

  \begin{equation}
    \label{eqn:acq_costs}
    \Lambda^{1st}(s) \defeq \left\{\begin{array}{lr}
                               \infty; & s > \bar{s}\\
                               0; & s < W(0)\\
                               sW^{-1}(s); & \mathrm{otherwise}
                             \end{array}\right.,
  \end{equation}

  is a proper\footnote{Recall that a function
    $f: \R \rightarrow (-\infty, \infty]$ is \textit{proper} if it is
    not everywhere equal to $+\infty$.}, lower semi-continuous,
  non-decreasing, and convex function on $\R$, which is strictly
  convex on $[W(0), \bar{s}]$.  Moreover, the convex conjugate

  \begin{equation*}
    \Lambda^*_{1st}(p) \defeq \underset{s \in \R}{\mathrm{sup}}\bigl[sp - \Lambda(s) \bigr],
  \end{equation*}

  is itself a proper, lower semi-continuous, non-decreasing, and
  convex function on $\R$, which satisfies $\Lambda^*_{1st}(p) = \infty$ for
  $p < 0$, and if $\bar{s} < \infty$, then
  $\Lambda^*_{1st}(p) = p\bar{x} - \Lambda_{1st}(\bar{s})$ on
  $p > \bar{x}$.
\end{theoremEnd}
\begin{proofEnd}
  Since $f(x) = xW(x)\ind_{\R_+}(x)$ we have, for $s \in \ran\ W$,
  $\Lambda(s) = s W^{-1}(s)$.  On $s \le W(0)$ we have
  $\Lambda(s) = 0$, and on $s \in \{s | s > \bar{s}\}$ we have
  $\Lambda(s) = \infty$.  Moreover, by Assumption
  \ref{ass:W_f_properties}, $W$ attains its supremum (when finite) and
  so $\Lambda$ is continuous at $\bar{s}$.  Convexity will therefore
  follow if $\Lambda$ is convex on $\intr\ \ran\ W$.  To this end, we
  use the $2$-concavity of $W$ to see that $1 - 1 / W(x)$ is concave
  on its domain and therefore that the inverse,
  $W^{-1}\bigl(1 / (1 - x)\bigr)$ is convex for the portion of
  $x \in (-\infty, 1)$ which remains in the range of $W$.  It is
  fairly well known that for a convex function $f$, the function

  \begin{equation*}
    (cx + d) f\bigl(\frac{ax + b}{cx + d}\bigr)
  \end{equation*}

  is convex on $cx + d > 0$ (see e.g. \cite[Ex. 3.20]{boyd2004convex}.
  Therefore, by setting $a = c = 1$, $b = -1$ and $d = 0$ we obtain
  convexity of

  \begin{equation*}
    (cs + d) W^{-1}\Bigl(\frac{cs + d}{(c - a) s + (d - b)}\Bigr) = sW^{-1}\bigl(s\bigr)
  \end{equation*}

  which is the function $\Lambda(s)$.

  In consideration of the conjugate, it is evident that since
  $\Lambda(s) \ge 0$ we have $\Lambda^*(p) = \infty$ for $p < 0$ (take
  $s \rightarrow -\infty$).  For $p > \bar{x}$, if $\bar{s} < \infty$,
  then since $s < \bar{s} \implies W^{-1}(s) < \bar{x}$ and
  $\Lambda(\bar{s}+) = \infty$ we see that $sp - \Lambda(s)$ is
  maximized at $s = \bar{s}$.  That $\Lambda^*(p)$ is non-decreasing
  follows since for $p \le q$ we have
  $sp - \Lambda(s) \le sq - \Lambda(s)$ on $s \ge 0$.  Finally, that
  $\Lambda^*(p)$ is proper, lower semi-continuous, and convex is a
  statement of the Fenchel-Moreau Theorem (see e.g., \cite[sec
  4.2]{clarke2013functional}).




\end{proofEnd}

\subsection{Optimal Bidding in Continuous Time}
\label{sec:optimal_bidding_in_ct}
As long as the acquisition cost functions are convex, we can
equivalently reformulate Problem \eqref{eqn:ct_problem} as the
following continuous time \textit{convex} optimization problem.  This
reformulation is not trivial and makes use of the existence of a
campaign-independent optimal bid $x_{ij}(t) = x_j(t)$, the existence
of which is established by using the convexity of $\Lambda$.

  \begin{theoremEnd}[end, restate, category=main, no link to proof]{proposition}[Continuous Primal Problem]
    \label{prop:ct_problem_cvx}
    In a first or second price auction, suppose that for each
    $j \in [M]$ and $t \in [0, T^j]$, the acquisition cost curve
    $\Lambda_j(x, t)$ is convex in $x$.  Then, Problem
    \eqref{eqn:ct_problem} can be reformulated as

    \begin{equation}
      \label{eqn:ct_problem_cvx}
      \begin{aligned}
        \underset{s, r}{\mathrm{minimize}} \quad & \sum_{j = 1}^M \int_0^{T^j}\Lambda_j(s_j(t), t) \d t\\
        \mathrm{subject\; to} \quad & \sum_{j \in \A_i}\int_0^{T_i}r_{ij}(t)\d t \ge C_i\\
        \quad & \sum_{i \in \B_j} r_{ij}(t) = s_j(t)\\
        \quad & r_{ij}(t) \ge 0.
      \end{aligned}
      \tag{$P$}
    \end{equation}

    If a solution exists, then a solution to the original problem
    \eqref{eqn:ct_problem} is obtained via
    $x_{ij}(t) = W_j^{-1}(s_j(t), t)$ for each $i \in \B_j$ and
    $\gamma_{ij}(t) = r_{ij}(t) / s_j(t)$.  Moreover, Problem
    \eqref{eqn:ct_problem_cvx} is convex.
  \end{theoremEnd}
  \begin{proofEnd}
    First, we establish that whenever a solution exists, there is also
    a solution with the property that $x_{ij}(t) = x_j(t)$ and where
    $\sum_{i \in \B_j} \gamma_{ij}(t) \in \{0, 1\}$.  To this end,
    suppose $(x, \gamma)$ is a solution of Problem
    \eqref{eqn:ct_problem}, and with total cost $J$. Let
    $(\tilde{x}, \tilde{\gamma})$ be another bid and allocation pair
    with total cost $\tilde{J}$ defined by
    \begin{align*}
      \tilde{x}_j(t)
      &\defeq W_j^{-1}\Bigl(\sum_{i \in \B_j}\gamma_{ij}(t)W_j(x_{ij}(t), t), t\Bigr),\\
      \tilde{\gamma}_{ij}(t)
      &\defeq \frac{\gamma_{ij}(t) W_j(x_{ij}(t), t)}{\sum_{u \in \B_j} \gamma_{uj}(t) W_j(x_{uj}(t), t)},
    \end{align*}
    where $0 / 0 \defeq 0$ in the definition of $\tilde{\gamma}$.  We
    proceed to show that $(\tilde{x}, \tilde{\gamma})$ is also a
    solution and we note that the definition of $\tilde{\gamma}$
    satisfies
    $\forall t < T^j\; \sum_{i \in \B_j}\tilde{\gamma}_{ij}(t) \in
    \{0, 1\}$.

    It is clear that the pair $\tilde{x}, \tilde{\gamma}(t)$ is
    feasible by construction.  Indeed, $\tilde{\gamma}_{ij}(t) \ge 0$ and
    $\sum_{i \in \B_j} \tilde{\gamma}_{ij}(t) \le 1$ by definition.
    Moreover, we have

    \begin{align*}
      \sum_{j \in \A_i} \tilde{\gamma}_{ij}(t)W_j(\tilde{x}_j(t), t)
      &= \sum_{j \in \A_i} \Bigl[\frac{\gamma_{ij}(t)W_j(x_{ij}(t), t) \sum_{v \in \B_j}\gamma_{vj}(t)W_v(x_{vj}(t), t)}{\sum_{u \in \B_j}\gamma_{uj}(t)W_j(x_{uj}(t), t)} \Bigr]\\
      &= \sum_{j \in \A_i} \gamma_{ij}(t)W_j(x_{ij}(t), t),
    \end{align*}

    which when integrated from $0$ to $T_i$ meets or exceeds $C_i$ since
    $(x, \gamma)$ is assumed to be a solution.
    
    The cost of $(\tilde{x}, \tilde{\gamma})$, instantaneously at time
    $t$, then satisfies $\tilde{J} = J$ since $J$ is the minimal cost and
    \begin{align*}
      \tilde{J}
      &\defeq \int_0^T \Bigl[\sum_{i = 1}^N \sum_{j \in \A_i} \tilde{\gamma}_{ij}(t) f_j(\tilde{x}_j(t)) \Bigr]\d t\\
      &\overset{(a)}{=} \int_0^T \Bigl[\sum_{i = 1}^N \sum_{j \in \A_i} \tilde{\gamma}_{ij}(t) \Lambda_j\Bigl(\sum_{u \in \B_j}\gamma_{uj}(t)W_j(x_{uj}(t), t), t\Bigr)\Bigr]\d t\\
      &\overset{(b)}{\le} \int_0^T \Bigl[\sum_{i = 1}^N \sum_{j \in \A_i} \tilde{\gamma}_{ij}(t) \sum_{u \in \B_j}\gamma_{uj}(t) \Lambda_j(W_j(x_{uj}(t), t), t)\Bigr]\d t\\
      &\overset{(c)}{=} \int_0^T \Bigl[\sum_{j = 1}^M\sum_{u \in \B_j} \gamma_{uj}(t) f_j(x_{uj}(t), t) \sum_{i \in \B_j} \tilde{\gamma}_{ij}(t)\Bigr]\d t = J\\
    \end{align*}

    where $(a)$ is just the definition of
    $\Lambda_j = f_j \circ W_j^{-1}$ (see Section
    \ref{sec:convex_acquisition_costs}), $(b)$ follows by the
    convexity of $\Lambda_j$ and that $\Lambda_j(0) = 0$ (since
    $\gamma_{ij}$ need not necessarily sum to $1$), and $(c)$ follows
    again by $\Lambda_j = f_j \circ W_j^{-1}$ and then by swapping the
    order of summation using $i \in \B_j \iff j \in \A_i$.

    We can now recall the original problem \eqref{eqn:ct_problem}, and
    apply the preceding results to eliminate the dependence of the bid
    on $i$, since if a solution exists it can be assumed to have the
    property $x_{ij}(t) = x_j(t)$:

    \begin{equation*}
      \begin{aligned}
        \underset{x, \gamma}{\text{minimize}} \quad & \sum_{i = 1}^N\int_{0}^{T_i} \Bigl[\sum_{j \in \A_i}\gamma_{ij}(t)f_j(x_j(t), t)\Bigr]\d t\\
        \textrm{subject to} \quad & \sum_{j \in \A_i}\int_0^{T_i}\gamma_{ij}(t)W_j(x_j(t), t) \d t\ge C_i\\
        \quad & \sum_{i \in \B_j} \gamma_{ij}(t) \le 1, \gamma_{ij}(t) \ge 0,
      \end{aligned}
    \end{equation*}

    Due to the bids' independence of $i$, we can rearrange the objective
    by swapping the order of summation:

    \begin{equation}
      \sum_{i = 1}^N\int_{0}^{T_i} \Bigl[\sum_{j \in \A_i}\gamma_{ij}(t)f_j(x_j(t), t)\Bigr]\d t
      = \sum_{j = 1}^M\int_0^{T^j} \Bigl[f_j(x_j(t), t) \sum_{i \in \B_j}\gamma_{ij}(t)\Bigr]\d t,
    \end{equation}

    which, after making the substitution $s_j(t) = W_j(x_j(t), t)$, results in
    \begin{equation}
      \begin{aligned}
        \underset{s, \gamma}{\text{minimize}} \quad & \sum_{j = 1}^M\int_{0}^{T^j} \Bigl[\Lambda_j(s_j(t), t) \sum_{i \in \B_j}\gamma_{ij}(t)\Bigr]\d t\\
        \textrm{subject to} \quad & \int_0^{T_i}\sum_{j \in \A_i}\gamma_{ij}(t)s_j(t) \ge C_i\\
        \quad & \sum_{i \in \B_j \cap \Tt} \gamma_{ij}(t) \le 1\\
        \quad & \gamma_{ij}(t) \ge 0, s_j(t) \ge 0.
      \end{aligned}
    \end{equation}

    Now, since any optimal solution can be assumed to be such that
    $\sum_{i \in \B_j}\gamma_{ij}(t) \in \{0, 1\}$ and since
    $\Lambda_j(0, t) = 0$, there is no loss of generality in adding
    the constraint $\sum_{i \in \B_j}\gamma_{ij}(t) = 1$ and
    eliminating this term from the objective.  Finally, making the
    substitution $r_{ij}(t) = \gamma_{ij}(t)s_j(t)$ results in the
    formulation of Problem \eqref{eqn:ct_problem_cvx}.
  \end{proofEnd}

  \begin{remark}[Connection to Transportation-Production Problems]
    This result demonstrates that Problem \eqref{eqn:ct_problem} can
    be reformulated as an instance of a (convex)
    \textit{Transportation-Production problem} (see
    \cite{leblanc1974transportation, sharp1970decomposition}) in
    continuous time (hence the label \eqref{eqn:ct_problem}), which is
    in turn an example of a \textit{monotropic program}
    \cite{rockafellar1984network, bertsekas2008extended}.  To see this
    connection, one can think of $i$ as the ``consumption nodes'', $j$
    as the ``production nodes'', the graph $\G$ (defined through the
    sets $\A_i, \B_j$) as defining the transportation network, and the
    functions $f_j, W_j, \Lambda_j$ as characterizing the costs of
    production.
  \end{remark}

  In order for a solution to exist, it is necessary only that an
  adequate amount of supply is available to fulfill each contract
  before its deadline.  In the context of internet advertising, this
  is not unreasonable, and a coarse sufficient condition is
  $W(x) \rightarrow \infty$ as $x \rightarrow \infty$.  As in the case
  of finite optimization problems, strong duality obtains if there is
  a \textit{strictly} feasible point.

  \begin{assumption}[Adequate Supply]
    \label{ass:adequate_supply}
    We suppose that $\exists x \in \R^M$ such that the following
    linear feasibility problem has a strictly feasible point (i.e., a
    Slater point)

    \begin{equation}
      \label{eqn:adequate_supply_problem}
      \begin{aligned}
        \mathrm{find} \quad \gamma_{ijk}\\
        \mathrm{such\ that} \quad & \sum_{k: T_k \le T_i} \sum_{j \in \A_i} \gamma_{ijk} \int_{T_{k - 1}}^{T_k} W_j(x_j, t)\d t > C_i\\
        \quad & \sum_{i \in \B_j}\gamma_{ijk} \le 1, \gamma_{ijk} \ge 0.\\
      \end{aligned}
    \end{equation}

    That is, there exists a large enough constant bid such that an
    adequate amount of supply is attained.
  \end{assumption}

  \begin{remark}
    A simpler sufficient condition (appearing also in
    \cite{leblanc1974transportation}) is to suppose that each
    curve has an adequate amount of supply to fulfill \textit{all} the
    contracts to which it is assigned:

    \begin{equation*}
      \forall j \in [M]\ \exists x \in \R:\ \int_0^{\tau_j} W_j(x, t)\d t > \sum_{i \in \B_j} C_i,
    \end{equation*}

    where $\tau_j = \text{min}\{T_i\ |\ i \in \B_j\}$.  Finally, this
    condition is clearly satisfied under the simple condition that if
    for each $j, t$ we have $W_j(x, t) \rightarrow \infty$ as
    $x \rightarrow \infty$.
  \end{remark}

  \begin{theoremEnd}[end, restate, category=main, no link to proof]{proposition}[Duality]
    \label{prop:ct_problem_dual}
    A Dual of \eqref{eqn:ct_problem_cvx} can be formulated as

    \begin{equation}
      \label{eqn:ct_problem_dual}
      \begin{aligned}
        \underset{\rho, \mu}{\mathrm{maximize}} \quad & -\sum_{j = 1}^M \sum_{k: T_k \le T^j} \int_{T_{k - 1}}^{T_k} \Lambda_j^*(\mu_{jk}, t) \d t + \sum_{i = 1}^N \rho_i C_i\\
        \mathrm{subject\; to} \quad & \mu_{jk} \ge \rho_i\ \forall i \in \B_j \cap \Tk\\
        \quad & \rho_i \ge 0,
      \end{aligned}
      \tag{$D$}
    \end{equation}

    which is a \textit{finite} convex problem.  Problem
    \eqref{eqn:ct_problem_dual} is dual to Problem
    \eqref{eqn:ct_problem_cvx} in the sense that if $D^\star$ and
    $P^\star$ are their respective values (possibly $\infty$ or
    $-\infty$), then $D^\star \le P^\star$.

    Finally, under Assumption \ref{ass:adequate_supply} there exists a
    solution $(s, r) \in L_2(\R)^{M} \times L_2(\R)^{N \times M}$ to
    Problem \eqref{eqn:ct_problem_cvx} and a solution
    $(\rho, \mu) \in \R^N \times \R^M$ to Problem
    \eqref{eqn:ct_problem_dual} and
    $-\infty < D^\star = P^\star < \infty$.

  \end{theoremEnd}

  \begin{proofEnd}
    The first part of the theorem is a simple statement of facts from
    convex analysis, and the strong duality result is based on
    Lagrange multiplier theory for real valued functions in a
    reflexive Banach space.  In particular, we establish the existence
    of a solution to Problem \eqref{eqn:ct_problem_cvx} (in $L_2$)
    under Assumption \ref{ass:adequate_supply} via the direct method
    (see e.g., \cite[thm. 5.51]{clarke2013functional}) and strong
    duality via the methods of \cite[Ch. 9]{clarke2013functional}.

    Consider a Lagrangian of Problem \eqref{eqn:ct_problem_cvx}:

    \begin{equation}
      \label{eqn:ct_lagrangian}
      \begin{aligned}
        \L(s, r, \rho, \mu)
        &= \sum_{j = 1}^M\int_0^{T^j}\Lambda_j(s_j(t), t)\d t + \sum_{i = 1}^N\rho_i \Bigl[C_i - \sum_{j \in \A_i}\int_0^{T_i}r_{ij}(t)\d t \Bigr]\\
        &\quad + \sum_{j = 1}^M  \int_0^{T^j} \mu_j(t)\Bigl[\sum_{i \in \B_j} r_{ij}(t) - s_j(t) \Bigr]\d t
      \end{aligned}
    \end{equation}

    with the problem domain $S = \text{dom }\mathcal{L}$ restricted to
    a convex subset of $L_2$:
    $$S \defeq \mathcal{P} \times \mathcal{D} \defeq \{(r, s), (\rho, \mu)\ |\ r_{ij}(t) \ge 0, \rho_i \ge 0,
    \mu_j(t) \ge 0\}.$$ It is well known that the primal problem
    \eqref{eqn:ct_problem_cvx} can be posed as finding $s, r$ which
    attains the infima:
    $$\underset{s, r \in \mathcal{P}}{\text{inf}}\ \underset{\rho, \mu \in \mathcal{D}}{\text{sup}}\ \L,$$ since the suprema is $\infty$ whenever the constraints of
    \eqref{eqn:ct_problem_cvx} are not met.  Moreover, it is a general
    theorem (the max-min inequality) that
    $$D^\star \defeq \underset{\rho, \mu \in \mathcal{D}}{\text{sup}}\ \underset{s,
      r \in \mathcal{P}}{\text{inf}}\ \L \le \underset{s, r \in \mathcal{P}}{\text{inf}}\
    \underset{\rho, \mu \in \mathcal{D}}{\text{sup}}\ \L \defeq P^\star.$$

    We proceed to show that finding $\rho, \mu$ to maximize
    $\inf_{s, r \in \mathcal{P}} \mathcal{L}$ is equivalent to
    \eqref{eqn:ct_problem_dual}.  Consider first the $r$ variable, we
    have:

    \begin{equation*}
      \underset{r \ge 0}{\text{inf}}\ \sum_{i = 1}^N\sum_{j \in \A_i}  \int_0^{T_i}\bigl(\mu_j(t) - \rho_i \bigr)r_{ij}(t) \d t,
    \end{equation*}

    which is $-\infty$ unless $\mu_j(t) \ge \rho_i$ for $i \in \B_j$.
    Next, examine $\text{inf}_s\ \mathcal{L}$:

    \begin{equation*}
      \underset{s}{\text{inf}}\ \sum_{j = 0}^M\int_0^{T^j} \Bigl[ \Lambda_j(s_j(t), t)\d t - \mu_j(t)s_j(t) \Bigr],
    \end{equation*}

    which can be computed point-wise (it will be clear that the
    resulting function is in $L_2$) as

    \begin{align*}
      \underset{s}{\text{inf}}\ \bigl[ \Lambda_j(s_j, t) - \mu_js_j \bigr]
      &= -\underset{s}{\text{sup}}\ \bigl[\mu_js_j - \Lambda_j(s_j, t) \bigr]\\
      &= -\Lambda_j^*(\mu_j, t),
    \end{align*}

    the negative of the conjugate of $s \mapsto \Lambda_j(s, t)$ (see
    Proposition \ref{prop:convex_acquisition_costs}).  Combining these
    calculations together results in the following problem:

    \begin{equation}
      \begin{aligned}
        \underset{\rho, \mu}{\mathrm{maximize}} \quad & -\sum_{j = 1}^M \int_0^{T^j} \Lambda_j^*(\mu_j(t), t) \d t + \sum_{i = 1}^N \rho_i C_i\\
        \mathrm{subject\; to} \quad & \mu_j(t) \ge \rho_i\ \forall i \in \B_j \cap \Tt\\
        \quad & \rho_i \ge 0.
      \end{aligned}
    \end{equation}

    Suppose that $\rho_i$ is optimal for this problem. Then, since
    $\Lambda^*$ is monotone increasing, the optimal $\mu_j(t)$ is the
    smallest possible, which implies we can take $\mu_j(t)$ to be
    piecewise constant between contract deadlines (any deviation over
    a positive measure interval increases the cost, and any deviation
    over a measure zero set doesn't change the cost).  Denote by
    $\mu_{jk}$ the value taken by the optimal $\mu_j(t)$ over
    $[T_{k - 1}, T_{k}]$ for $T_k \le T^j$ to obtain Problem
    \eqref{eqn:ct_problem_dual}.  Indeed, it can be seen that
    $\mu_{jk} = \underset{i \in \B_j \cap \Tk}{\text{max}}\ \rho_i$.

    Finally, we show the existence of solutions to both
    \eqref{eqn:ct_problem_cvx} and \eqref{eqn:ct_problem_dual}, as
    well as the equality of their values, under the adequate supply
    assumption \ref{ass:adequate_supply}.

    Under Assumption \ref{ass:adequate_supply} the $\R$-valued
    function on $L_2(\R)^M \times L_2(\R)^{N \times M}$ defined by
    $(s, r) \mapsto \underset{\rho, \mu}{\text{sup}}\ \L(s, r, \rho,
    \mu)$ is a proper, convex, and l.s.c. function.  That this
    function is proper follows from the assumption since there is at
    least one feasible point, the convexity follows since suprema over
    families of affine functions are convex, and lower semi-continuity
    follows since this property is preserved through the suprema.
    Naturally, the restriction of $S$ to the primal variables $(s, r)$
    remains closed and convex.

    Now, by the basic properties of $W_j$ (Assumption
    \ref{ass:W_f_properties}) $\Lambda_j(s)$ is positively-coercive
    (i.e.  $\Lambda_j(s) \rightarrow \infty$ as
    $s \rightarrow \infty$); this property is thus inherited by the
    function
    $s \mapsto \underset{\rho, \mu}{\text{sup}}\ \L(s, r, \rho, \mu).$
    Indeed, by the construction of $\L$ the function
    $r \mapsto \underset{\rho, \mu}{\text{sup}}\ \L(s, r, \rho, \mu)$
    must be positively-coercive as well since if
    $\sum_{i \in \B_j}r_{ij}(t) \ne s_j(t)$ then
    ${\text{sup}}\ \L(s, r, \rho, \mu) = \infty$.  Then, keeping in
    mind that $(s, r) \in S \implies s, r \ge 0$, this is enough to
    satisfy the growth condition required by \cite[thm. 5.51
    (c)]{clarke2013functional} and establishes the existence of a
    solution to
    $\underset{s, r}{\text{inf}}\ \underset{\rho, \mu}{\text{sup}}\
    \L$ and therefore to $(P)$.

    The statement of strong duality now follows from combining
    \cite[thm. 9.8]{clarke2013functional} and
    \cite[thm. 9.13]{clarke2013functional} with Assumption
    \ref{ass:adequate_supply}.
\end{proofEnd}

Finally, the optimal bid path $x(t)$ for Problem
\eqref{eqn:ct_problem} can be reconstructed from solutions to
\eqref{eqn:ct_problem_dual} and is, in fact, fully characterized by
the finite vector $\rho \in \R^N$, which we designate as the vector of
\textit{pseudo-bids}.

\begin{theoremEnd}[normal]{example}[Pseudo Bids]
  \label{ex:pseudo_bids}
  Suppose that $\mu_{jk}, \rho_i$ are optimal dual multipliers for the
  Problem \eqref{eqn:ct_problem_dual} and consider the $s$-dependent
  portion of the Lagrangian:

  \begin{equation*}
    \begin{aligned}
      L(s) = \sum_{j = 1}^M\int_0^{T^j}\Bigl[\Lambda_j(s_j(t), t) - \mu_j(t)s_j(t) \Bigr]\d t,
    \end{aligned}
  \end{equation*}

  where $\mu_j(t) = \mu_{jk}$ for $t \in [T_{k - 1}, T_k)$ is
  piecewise constant.

  Minimizing this function over $s_j(t)$ leads to an optimal supply
  path as a function of $\mu_j(t)$.  Since the above function is
  convex, we know that the optimum is attained at $s_j(t)$ such that
  the pointwise selection holds:

  \begin{equation}
    \label{eqn:pseudo_bid_equation}
    \mu_j(t) \in \partial \Lambda_j(s_j(t), t),
  \end{equation}

  where $\partial$ indicates the subdifferential and is with respect
  to the first argument.  Therefore, since
  $\mu_j(t) = \underset{i \in \B_j \cap \Tt}{\text{max}}\ \rho_i$, the
  entire optimal supply path $s(t)$, and therefore the entire optimal
  bid path $x(t)$, is fully determined by the \textit{finite} vector
  $\rho \in \R^N$ through Equation \eqref{eqn:pseudo_bid_equation}.
  For this reason, we refer to $\rho$ as the vector of \textit{pseudo
    bids}.
\end{theoremEnd}

\subsection{Connections Between First and Second Price Auctions}
\label{sec:connections}
It will be seen in Section \ref{sec:simulation} that, empirically, the
optimal solution calculated for the second price auction also provides
very good bid and allocation paths for the first price auction.  From
Example \ref{ex:pseudo_bids}, we know that the optimal solutions
cannot be exactly the same (except perhaps in some coincidental
cases), but we explain the empirical similarity by illuminating the
close connection between the first and second price cases in the
following proposition.

\begin{theoremEnd}[restate, end, category=aux, no link to proof]{proposition}[Connection Between First and Second Price Case]
  \label{prop:connection}
  Under Assumption \ref{ass:adequate_supply}, and whenever the
  acquisition cost curves are convex, the Problem
  \eqref{eqn:ct_problem} for a second price auction is equivalent to
  solving

  \begin{equation}
    \begin{aligned}
      \underset{x, \gamma, \rho}{\mathrm{minimize}} \quad & \sum_{j = 1}^M\int_0^{T^j} x_j(t)W_j(x_j(t), t) \d t - \sum_{i = 1}^N \rho_i C_i\\
      \mathrm{subject\ to} \quad & \sum_{j \in \A_i}\int_0^{T_i}\gamma_{ij}(t)W_j(x_j(t), t) \ge C_i\\
      \quad & \sum_{i \in \B_j} \gamma_{ij}(t) = 1, \gamma_{ij}(t) \ge 0, \rho_i \ge 0\\
      \quad & x_j(t) \ge \rho_i\ \forall i \in \B_j \cap \Tt.
    \end{aligned}
  \end{equation}

  The objective function involves $x_j(t)W_j(x_j(t), t)$, which is the
  cost associated with a first price auction.
\end{theoremEnd}
\begin{proofEnd}
  We first examine the conjugate function for the second price
  acquisition function.  We have

  \begin{equation}
    \Lambda_{2nd}(s) \defeq \left\{
      \begin{array}{lr}
        \infty; & s > \bar{s}\\
        0; & s < W(0)\\
        \int_{W(0)}^s W^{-1}(y)\d y; & \mathrm{otherwise}
      \end{array}\right..
  \end{equation}

  We can find the conjugate on a segment of $\R$ by differentiation:

\begin{equation*}
  \Lambda^*_{2nd}(\mu) = \underset{s}{\text{sup}}\bigl[s\mu - \Lambda(s) \bigr],
\end{equation*}

leading to the supremum being attained at $s = W(\mu)$ as long as
$\mu \in [0, \bar{x}]$ (in which case the derivative is indeed
$W^{-1}(s)$).  We already know from Proposition
\ref{prop:convex_acquisition_costs} that when $\mu < 0$ we have
$\Lambda^*(\mu) = \infty$ and if $\bar{s} < \infty$ and
$\mu > \bar{x}$, then
$\Lambda^*(\mu) = \bar{s}\mu - \Lambda(\bar{s})$.

Therefore, since $\Lambda(W(\mu)) = f(\mu)$ by definition, we have

\begin{equation}
  \label{eqn:2pa_conjugate}
  \Lambda^*_{2nd}(\mu) = \left\{
    \begin{array}{lr}
      \infty; & \mu < 0\\
      \mu \bar{s} - \Lambda_{2nd}(\bar{s}); & \mu > \bar{x}\\
      \mu W(\mu) - f_{2nd}(\mu); & \text{otherwise}
    \end{array}\right..
\end{equation}

Given a dual solution $\mu, \rho$ of Problem
\eqref{eqn:ct_problem_dual}, which attains the primal problem's value
by strong duality Proposition \ref{prop:ct_problem_dual}, we see from
the above calculations that the maximizing argument for $\Lambda^*$ is
just $s = W(\mu)$, therefore it must be the case that the optimal bid
is the optimal multiplier $x_j(t) = \mu_{jk}$ on
$t \in [T_{k - 1}, T_k)$, which recovers the result of Example
\ref{ex:pseudo_bids} that for the second price auction there exists an
optimal piecewise constant bid.

Using this fact, and writing the dual as a minimization problem by
negating its objective, we can consider the problem of
\textit{simultaneously} solving the dual \eqref{eqn:ct_problem_dual}
and the original Problem \eqref{eqn:ct_problem}

\begin{equation}
  \begin{aligned}
    \underset{x, \gamma, \rho}{\text{minimize}} \quad & \sum_{j = 1}^M\int_0^{T^j} \bigl[f_j^{2nd}(x_j(t), t) + \Lambda_j^*(x_j(t), t)\bigr] \d t - \sum_{i = 1}^N \rho_i C_i\\
    \textrm{subject to} \quad & \sum_{j \in \A_i}\int_0^{T_i}\gamma_{ij}(t)W_j(x_j(t), t) \ge C_i\\
    \quad & \sum_{i \in \B_j} \gamma_{ij}(t) = 1, \gamma_{ij}(t) \ge 0, \rho_i \ge 0\\
    \quad & x_j(t) \ge \rho_i\ \forall i \in \B_j \cap \Tt,
  \end{aligned}
\end{equation}

where we have again applied the fact that there exists a solution such
that $x_{ij}(t) = x_j(t)$ for every $i \in \B_j$ and where
$\sum_{i \in \B_j}\gamma_{ij}(t) = 1$ (see the proof of
\ref{prop:ct_problem_cvx}).

If we can show that the optimal dual multipliers satisfy
$\mu_{jk} \le \bar{x}_j$, then using the fact that $x_j(t) = \mu_{jk}$
for $t \in [T_{k - 1}, T_k)$ and the expression for $\Lambda^*$ we
have the result, since the $f_j^{2nd}$ terms cancel.

To this end, recognize that since
$\mu_{jk} = \underset{i \in \B_j \cap \Tk}{\text{max}}\ \rho_i$, we
must have $\mu_{jk} \le \bar{x}_j$ unless there is some
$\rho_i > \bar{x}_j$.  We suppose by way of contradiction that
$\rho_i > \bar{x}_j$.  Consider the portion of the objective of
\eqref{eqn:ct_problem_dual} related to contract $i$ (where
$\Lambda = \Lambda_{2nd}$):

\begin{align*}
  -\sum_{j \in \A_i}\sum_{k: T_k \le T_i}\int_{T_{k - 1}}^{T_k}\Lambda_j^*(\mu_{jk}, t)\d t + \rho_iC_i
  &\overset{(a)}{\le} \rho_iC_i - \sum_{j \in \A_i}\int_0^{T_i}\Lambda_j^*(\rho_i, t)\d t\\
  &\overset{(b)}{=} \rho_iC_i - \sum_{j \in \A_i}\int_0^{T_i}(\rho_i \bar{s}_j(t) - \Lambda_j(\bar{s}_j(t), t))\d t\\
  &= \sum_{j \in \A_i}\int_0^{T_i}\Lambda_j(\bar{s}_j, t)\d t + \rho_i\Bigl[C_i - \int_0^{T_i}\bar{s}_j(t)\d t \Bigr]\\
  &\overset{(c)}{<} \sum_{j \in \A_i} \int_0^{T_i} f_j(\bar{x}_j(t), t)\d t
\end{align*}

where $(a)$ follows from $\mu_{jk} \ge \rho_i$ since $\Lambda^*$ is
non-decreasing, $(b)$ from Equation \eqref{eqn:2pa_conjugate} since
$\rho_i > \bar{x}_j$, and $(c)$ from $\rho_i \ge 0$ combined with
Assumption \ref{ass:adequate_supply} which ensures the bracketed term
is negative.  The final expression is the cost attained by bidding
$\bar{x}_j(t)$ so by strong duality $\rho_i > \bar{x}_j$ must not have
been optimal (recall that in the dual we are maximizing).  Therefore,
$\mu_{jk} \le \bar{x}_j$, which completes the proof.
\end{proofEnd}

Returning to Example \ref{ex:pseudo_bids}, we can make a further
explicit comparison between the two cases.

\begin{theoremEnd}[normal]{example}[Pseudo Bids (continued)]
  \label{ex:pseudo_bids2}
  For the purpose of illustration, let us temporarily assume that our
  supply rate curves are smooth in the bid $x$, and that the optimal
  solutions satisfy $s_j(t) \in (W_j(0, t), \bar{s}_j(t))$ (i.e., do
  not reach the extremes of available supply) so that we have the
  simple expressions
  $\Lambda^{2nd}_j(s, t) = \int_{W_j(0, t)}^s W_j^{-1}(y, t)\d y$ and
  $\Lambda^{1st}_j(s) = sW_j^{-1}(s, t)$.  For the first price case,
  let us define a function $g_j$, where $'$ denotes differentiation
  with respect to $x$:

  \begin{equation*}
    g_j(x, t) \defeq \frac{f_{j, 1st}'(x, t)}{W_j'(x, t)} = x + \frac{W_j(x, t)}{W_j'(x, t)}.
  \end{equation*}

  We can then solve Equation \ref{eqn:pseudo_bid_equation} for the
  optimal bid in terms of $\mu_j$, which in the first price case, can
  be written in terms of $g$:
  $\Lambda'^{-1}_j(x, t) = W_j \circ g_j^{-1}(x, t)$.  The second
  price case is much simpler, and we have
  $\Lambda'_{2nd}(x, t) = W^{-1}(s, t)$.  That is, the analogous '$g$'
  function for the second price case is the identity.

  Now, in each case we map the desired supply $s$ into the required
  bid $x$ through $x_j(t) = W_j^{-1}(s_j(t), t)$ and find that the
  optimal bids for each case are

  \begin{equation}
    x_j(t) = \left\{
      \begin{array}{lr}
        \mu_j^{2nd}(t) & \text{Second Price}\\
        g_j^{-1}(\mu_j^{1st}(t), t) & \text{First Price},
      \end{array}\right.
  \end{equation}

  where $\mu_j^{2nd}(t)$ is the dual multipliers associated to Problem
  \eqref{eqn:ct_problem_dual} with second price cost function and
  $\mu_j^{1st}$ are multipliers of the same problem but with the first
  price cost function; that is, $\mu_j^{1st} \ne \mu_j^{2nd}$ in
  general.  It is important to recognize that $\mu(t)$ is a piecewise
  constant function which is fully determined by $\rho \in \R^N$.  The
  implication being that in each case, the entire time-dependent
  optimal bid path is fully characterized by the finite vector $\rho$
  associated to the dual problem \eqref{eqn:ct_problem_dual}.

  The upshot of these calculations is that in the second price case,
  the optimal bid path is constant between contract deadlines, whereas
  in the first price case the pseudo bids need to be mapped through
  the function $g^{-1}$.  The intuition for this fact is that in a
  second price case you are ``protected'' from overpaying (since you
  only pay the second highest bid) whereas in the first price case you
  need to adapt the bids to your belief about the prevailing prices.

  If the supply rate curves are time-independent (i.e.
  $W(x, t) = W(x)$), then the function $g_j^{-1}$ is time-independent
  as well.  It may be conjectured that in this case the optimal bids
  would coincide, i.e., that
  $\mu_j^{2nd}(t) = g_j^{-1}(\mu_j^{1st}(t))$.  Experimental evidence
  shows that this does not seem to be the case in general.
\end{theoremEnd}

\section{Computational Methods and Simulations}
\label{sec:computational_methods}
In this section we review computational methods useful for the
implementation of our theory (Section \ref{sec:approximation}); carry
out a Monte Carlo simulation lending evidence to the similarity
between the first and second price case, as suggested by Proposition
\ref{prop:connection} (Section \ref{sec:simulation}); and finally
provide basic results on practical methods for adapting to a
stochastic environment (Section \ref{sec:improving_fulfillment}).

\subsection{Computing and Implementing Optimal Bids}
\label{sec:approximation}
To summarize the developments so far, the solution to the optimal
bidding and allocation problem \eqref{eqn:ct_problem} consists of a
bid path $x(t)$ and allocation path $\gamma(t)$.  The function
$\gamma_{ij}(t)$ indicates that if an item of type $j$ is won at
auction at time $t$, then it should be allocated towards contract $i$
with probability $\gamma_{ij}(t)$.  Therefore, upon the arrival of an
item of type $j$, the bidder should sample a random index
$\hat{i} \in \B_j$ from a categorical distribution with
$\P\{\hat{i} = i\} = \gamma_{ij}(t)$.  Then, the function
$x_{\hat{i}j}(t)$ indicates the nominal bid which should be submit to
the auction house (the actual bid being
$\mathcal{N}(x_{\hat{i}j}(t), \sigma^2)$ if randomization is used).
Finally, if the item is won, it should be allocated to contract
$\hat{i}$.

The Problem \eqref{eqn:ct_problem} is, as written, not tractable.
However, Proposition \ref{prop:ct_problem_cvx} tells us that the
solution $s(t), r(t)$ to the \textit{convex} problem
\eqref{eqn:ct_problem_cvx} can be used to obtain the optimal bid and
allocation through

\begin{align*}
  x_{ij}(t) &= W_j^{-1}(s_j(t), t),\\
  \gamma_{ij}(t) &= r_{ij}(t) / s_j(t),
\end{align*}

where it is to be understood that $x_{ij}(t) = \gamma_{ij}(t) = 0$ for
any $j \not \in \A_i$ or $t > T_i$, and that any $0 / 0$ encountered
in the construction of $\gamma_{ij}(t)$ is to be taken as $0$.
Moreover, this formula tells us that we actually need to keep track of
only $M$ bidding functions, $x_j(t)$, since there is an optimal
solution such that $x_{ij}(t) = x_{i'j}(t)$ for each $i, i' \in \B_j$.

Problem \eqref{eqn:ct_problem_cvx} is still an infinite dimensional
problem, and thus remains intractable as written.  A solution could be
obtained by first solving the dual \eqref{eqn:ct_problem_dual} (which
is finite, but requires integration), substituting the resulting
pseudo-bids (through $g^{-1}$, see Example \ref{ex:pseudo_bids2}) into
the original problem \eqref{eqn:ct_problem} and solving the resulting
linear program to obtain $\gamma(t)$.  However, we have found that
directly discretizing the primal problem and solving the resulting
convex program tends to be faster and with limited degradation in
accuracy.  Moreover, solving the primal directly results in
\textit{both} the optimal bids and the optimal allocations
simultaneously.

Therefore, in this section we study the discretization of Problem
\eqref{eqn:ct_problem_cvx}.  We will see that the linearity of the
constraints allows us to construct \textit{feasible} approximate
solutions for \eqref{eqn:ct_problem_cvx}, and that we can also
estimate the sub-optimality of the approximation.  Some additional
details pertaining to the convexity of $\Lambda$ and how the supply
curves can be guaranteed to have the needed $\alpha$-concave property
are provided in the Appendix \ref{sec:supply_curve_estimation}.  The
conclusion of our developments is provided in Algorithm
\ref{alg:eli5_bidding}.  The $\texttt{error}$ condition in Algorithm
\ref{alg:eli5_bidding} encountered when supply is inadequate can be
handled, for example, by modifying the objective of Problem
\eqref{eqn:dt_problem_cvx} to instead penalize supply shortfalls,
rather than attempting to enforce them as constraints, resulting in a
``best effort'' solution.

To the end of discretizing Problem \eqref{eqn:ct_problem_cvx}, choose
a sequence of points $0 = \Tb_0 < \Tb_1 < \cdots < \Tb_K = T$ where
$K \ge N$ and $\{T_i\}_{i = 1}^N \subseteq \{\Tb_k\}_{k = 1}^K$.
Then, using the notation $\Delta_k = \Tb_{k} - \Tb_{k - 1}$ we define,
for $k \in [K]$ and $\Tb_k \le T^j$

\begin{equation}
  \label{eqn:lambda_bar}
  \Lb_{jk}(s) \defeq \frac{1}{2}\Delta_k\bigl[\Lambda_j(s, \Tb_{k}) + \Lambda_j(s, \Tb_{k - 1}) \bigr],
\end{equation}

i.e., a trapezoidal approximation of the integral
$\int_{\Tb_{k - 1}}^{\Tb_k} \Lambda_j(s, t) \dt$.  The following finite
approximation of Problem \ref{eqn:ct_problem_cvx} is natural:

\begin{theoremEnd}[restate, end, category=aux2, no link to proof]{proposition}[Finite Primal Problem]
  \label{prop:finite_primal}
  The finite optimization problem over the variables
  $r_{ij}[k], s_{ij}[k]$ defined by

  \begin{equation}
    \label{eqn:dt_problem_cvx}
    \begin{aligned}
      \underset{s, r}{\mathrm{minimize}} \quad & \sum_{j = 1}^M \sum_{k: \Tb_k \le T^j} \Lb_{jk}(s_j[k])\\
      \mathrm{subject\; to} \quad & \sum_{j \in \A_i} \sum_{k: \Tb_k \le T_i} \Delta_k r_{ij}[k] \ge C_i\\
      \quad & \sum_{i \in \B_j \cap \Tk} r_{ij}[k] = s_j[k]\\
      \quad & r_{ij}[k] \ge 0,
    \end{aligned}
    \tag{$\PK{K}$}
  \end{equation}

  is a finite approximation of \eqref{eqn:ct_problem_cvx}: any
  solution $(s_j[k], r_{ij}[k])$ of \eqref{eqn:dt_problem_cvx} is
  feasible for Problem \eqref{eqn:ct_problem_cvx} (as a piecewise
  constant function of $t$).

  Let $g_j^{-1}(\mu, t)$ denote a continuous selection from Equation
  \eqref{eqn:pseudo_bid_equation} of Example \ref{ex:pseudo_bids}
  (e.g., choose the right derivative).  If each cost function
  $f_j(x, t) = xW_j(x, t)$ is $L_f$-Lipschitz in $x$ (uniformly in $t$
  a.e.) and $g_j^{-1}(x, t)$ is $\Gamma_g$-Lipschitz in $t$
  a.e. (uniformly in $x$), then the cost difference $\epsilon_K$ of
  the optimal solution to \eqref{eqn:ct_problem_cvx} and the feasible
  piecewise constant approximation obtained from Problem
  \eqref{eqn:dt_problem_cvx} is bounded by

  \begin{equation*}
    \epsilon_K \le \frac{M}{4}L_f\Gamma_g \sum_{k = 1}^K \Delta_k^2.
  \end{equation*}

  If for each $j$ the function $W_j(x, t)$ (and hence
  $\Lambda_j(s, t)$) is almost everywhere twice continuously
  differentiable in $t$, the integral approximation error $\delta_K$
  between the objective of \eqref{eqn:ct_problem_cvx} and
  \eqref{eqn:dt_problem_cvx} is bounded by
  $\delta_K = \mathcal{O}(\sum_{k = 1}^K \Delta_k^3)$.

  In particular, if we have $\Delta_k = T / K$ (valid as long as
  $\Tb_k$ contains the contract deadlines), then we have the bounds

  \begin{align*}
    \epsilon_K &\le \frac{ML_f \Gamma_gT^2}{4K}\\
    \delta_K &= \mathcal{O}(1/K^2).
  \end{align*}

  The total discretization error between the value of
  \eqref{eqn:dt_problem_cvx} and the value of
  \eqref{eqn:ct_problem_cvx} is therefore
  $\epsilon_K + \delta_K = \mathcal{O}(1/K)$.
\end{theoremEnd}
\begin{proofEnd}
  Let $(s^\star[k], r^\star[k])$ be an optimal solution to Problem
  \eqref{eqn:dt_problem_cvx} and $(s^\star(t), r^\star(t)$ an optimal
  solution to \eqref{eqn:ct_problem_cvx} with corresponding (piecewise
  constant) dual multipliers $\mu_j^\star(t)$.  It is evident that
  $(s^\star[k], r^\star[k])$ is feasible for the continuous time
  problem \eqref{eqn:ct_problem_cvx} by construction since the
  integral of a piecewise constant function is e.g.,
  $\Delta_k r_{ij}[k]$.

  Bounding the integral approximation error is a simple application of
  the well known error bound for the trapezoidal rule (see
  e.g. \cite[ch. 7]{owen2013monte}), i.e., for any point $s_0$
  there is some $\hat{t} \in (\Tb_{k - 1}, \Tb_k)$ such that
  $$\bigl|\int_{\Tb_{k - 1}}^{\Tb_k} \Lambda_j(s_0, t) \d t\bigr|
  \le \frac{\Delta_k^3}{24}\frac{\partial^2\Lambda_j(s_0,
    t)}{\partial t^2}\Bigr|_{t = \hat{t}}.$$

  Now the integral approximation error, i.e., the error between the
  objective of \eqref{eqn:dt_problem_cvx} and
  \eqref{eqn:ct_problem_cvx} evaluated on the piecewise constant
  solution $s^\star[k]$ of \eqref{eqn:dt_problem_cvx} is

  \begin{equation*}
    \Bigl|\sum_{j = 1}^M \sum_{k: \Tb_k \le T^j}\int_{\Tb_{k - 1}}^{\Tb_k} \Lambda_j(s^\star[k], t)\d t
    - \sum_{j = 1}^M\sum_{k: \Tb_k \le T^j} \Lb_{jk}(s^\star(\Tb_k))\Bigr| \le \mathcal{O}\Bigl(M\sum_{k = 1}^K\Delta_k^3\Bigr).
  \end{equation*}

  To bound the error of using the solution to
  \eqref{eqn:dt_problem_cvx} as a feasible point for
  \eqref{eqn:ct_problem_cvx}, we would like to find a point in time
  where attained supply is maximized:

  \begin{equation*}
    \tau_k \in \Bigl\{t \in [\Tb_{k - 1}, \Tb_k)\ |\ s^\star(\tau_k) \ge s^\star(u)\ \forall u \in [\Tb_{k - 1}, \Tb_k] \Bigr\},
  \end{equation*}

  though this set may be empty since $s^\star(t)$ can be discontinuous
  at $\Tb_k$.  However, since we are here assuming smoothness of the
  supply rate curves in $t$, the total supply attained in an
  $\epsilon$-interval of time must converge to $0$ as
  $\epsilon \rightarrow 0$ and therefore there must exist points
  $\tau_k \in [\Tb_{k - 1}, \Tb_k)$ such that the piecewise constant
  function $k \mapsto s^\star(\tau_k)$ is feasible for
  \eqref{eqn:dt_problem_cvx}, since $s^\star(t)$ is feasible for
  \eqref{eqn:ct_problem_cvx}.  We therefore have the inequalities:

  \begin{equation*}
    \int_{\Tb_{k - 1}}^{\Tb_k}\Lambda_j(s^\star_j(t), t)\d t
    \le \int_{\Tb_{k - 1}}^{\Tb_{k}}\Lambda_j(s^\star_j[k], t)\d t
    \le \int_{\Tb_{k - 1}}^{\Tb_{k}}\Lambda_j(s^\star_j(\tau_k), t)\d t.
  \end{equation*}

  We can bound the absolute difference between the outer terms as follows:

  \begin{align*}
    \int_{\Tb_{k - 1}}^{\Tb_k}|\Lambda_j(s^\star_j(t), t) - \Lambda_j(s^\star_j(\tau_k), t)|\d t
    &\overset{(a)}= \int_{\Tb_{k - 1}}^{\Tb_k}|f_j(g^{-1}_j(\mu^\star_j(t), t), t) - f_j(g^{-1}_j(\mu^\star_j(t), \tau_k), t)|\d t\\
    &\overset{(b)}\le L_f \int_{\Tb_{k - 1}}^{\Tb_k}|g^{-1}_j(\mu^\star_j(t), t) - g^{-1}_j(\mu^\star_j(t), \tau_k)|\d t\\
    &\overset{(c)}\le L_f\Gamma_g \int_{\Tb_{k - 1}}^{\Tb_k}|t - \tau_k|\d t\\
    &\le \frac{1}{4}L_f\Gamma_g \Delta_k^2,
  \end{align*}

  where $(a)$ follows by definition of $\Lambda_j$ and $g^{-1}$, $(b)$
  follows by the Lipschitz continuity of $f(x, t)$ w.r.t. $x$; and
  $(c)$ follows by the Lipschitz continuity of $g^{-1}(x, t)$
  w.r.t. $t$.

  Summing these individual error terms we obtain an overall error
  bound

  \begin{equation*}
    \sum_{j = 1}^M\Bigl|\int_0^{T}\Lambda(s^\star(t), t)\d t - \sum_{k = 1}^K \Delta_k \Lb_k(s^\star[k])\Bigr| \le \frac{M}{4}L_f\Gamma_g \sum_{k = 1}^K \Delta_k^2.
  \end{equation*}

  The total error then arises from summation

  \begin{align*}
    \Bigl|\int_{\Tb_{k - 1}}^{\Tb_k} \Lambda_j(s^\star(t))\d t - \Lb_j(s^\star[k])\Bigr|
    &\le \Bigl|\int_{\Tb_{k - 1}}^{\Tb_k} \Lambda_j(s^\star(t))\d t - \int_{\Tb_{k - 1}}^{\Tb_k}\Lambda_j(s^\star(\Tb_k))\Bigr|\\
    &\quad + \Bigl|\int_{\Tb_{k - 1}}^{\Tb_k}\Lambda_j(s^\star(\Tb_kk)) - \Lb_j(s^\star(\Tb_k))\Bigr|,
  \end{align*}

  and results in a total error $\epsilon_K + \delta_K = \mathcal{O}(1/K)$.
\end{proofEnd}

A summary of our proposed bidding methods are provided in Algorithm
\ref{alg:eli5_bidding}.  The cost functions must correspond to either
a second or a first price auction, and the supply curves must be
strictly $2-$concave in the latter case, see Proposition
\ref{prop:convex_acquisition_costs}.

\begin{algorithm}
  \SetKwInOut{Input}{input}
  \SetKwInOut{Output}{output}
  \SetKwInOut{Initialize}{initialize}
  \DontPrintSemicolon

  \BlankLine
  \caption{Computing Optimal Bids}
  \label{alg:eli5_bidding}

  \Input{Contracts $\{(\A_i, C_i, T_i)\}_{i = 1}^N$, supply curves
    $\{W_j(x, t)\}_{j = 1}^M$, cost functions $f_j(x, t)$, parameter $K \ge N + 1$}

  \Output{Bid path $x(t)$ and allocation path $\gamma(t)$}

  \BlankLine

  $(\Tb_k)_{k = 0}^{K - N} \leftarrow \texttt{segment}([0, T], K - N)$ \texttt{//  cut $[0, T]$ into $K - N$ equal segments}\\
  $(\Tb_k)_{k = 0}^K \leftarrow \texttt{sort}(\{\Tb_k\}_{k = 0}^{K - N} \cup \{T_i\}_{i = 1}^N)$ \texttt{// incorporate contract deadlines}\\
  $\texttt{Let } \Lambda_{j}(s, t) = f_j \circ W_j^{-1}(s, t)$ \texttt{//  acquisition function}\\
  \If{Assumption \ref{ass:adequate_supply} holds (adequate supply is available)}{
    $s_j[k], r_{ij}[k] \leftarrow \texttt{solve}(P_K)$ \texttt{//  solve discretized problem}\\
  }
  \Else{
    \Return \texttt{error}
  }
  $x_j(t) \leftarrow W_j^{-1}(s_j[k], t)\ \forall t \in [\Tb_{k - 1}, \Tb_k), k: \Tb_k \le T^j, j \in [M]$ \texttt{// construct bid path}\\
  $\gamma_{ij}(t) \leftarrow r_{ij}[k] / s_j[k]\ \forall t \in [\Tb_{k - 1}, \Tb_k), k: \Tb_k \le T^j, i \in \B_j, j \in [M]$\\
  \Return $(x(t), \gamma(t))$
\end{algorithm}

\subsection{Monte Carlo Simulations}
\label{sec:simulation}
We have run experiments to empirically evaluate the performance of our
methods on real data from the well known IPinYou dataset
\cite{liao2014ipinyou, zhang2014real}.  Further details on the
specific simulation methodology are provided in the appendix.  Our
simulations in this section focus on the similarity between the
optimal solutions computed in the first and second price settings
cf. Proposition \ref{prop:connection}.

All our computations have been carried out with Python's scientific
computing stack \cite{2020SciPy-NMeth} and CVXPY
\cite{diamond2016cvxpy} \cite{agrawal2018rewriting}.  To summarize our
setup, we have estimated supply rate curves $W_j(x, t)$ via Gaussian
kernel density estimation using the one week of available data
stratified into twenty-four one hour intervals.  The bandwidth is
chosen via Silverman's rule (see e.g.,
\cite[ch. 6]{wasserman2006all}), which results in smooth estimates,
and is not likely to be overfit.  The hourly stratification results in
a supply rate curve estimate which accounts for daily periodic trends,
but not weekly ones.  In practice, supply curve estimation is made
more challenging by the fact that price observations are
\textit{censored}, that is, the true winning price is only observed by
the winner \cite{wu2015predicting, zhang2016bid, wu2018deep}.
However, the IPinYou dataset was obtained by submitting very high bids
which win almost every impression, and we are therefore able to
comfortably ignore the affects of censoring in our simulation with the
understanding that more sophisticated supply rate curve estimation
methods are to be applied in practice.

The problems we simulate in this section contain $N = 6$ contracts and
$M = 5$ item types.  Each contract in the collection is randomly
sampled uniformly at random within prescribed bounds.  In particular,
the starting ``real'' time point (i.e., time $T_0 = 0$) is sampled
from anywhere between the bounds of available data, the length of
contracts is uniformly random between $0$ and $70$ hours, and the
number of required items are sampled uniformly between a small (easily
fulfilled) lower bound, and a large (near the maximum available
supply) upper bound -- random contracts which are not feasible are
re-sampled.  A total of 250 Monte Carlo iterations are carried out.
We use the same estimated supply rate curves in each simulation, but
the auction prices and item arrival times are sampled directly from
real data\footnote{All of the code used in our experiments will be made
available on github \textsf{@RJTK}.}.

Whenever an adequate number of items to fulfill a contract are
acquired, that contract is removed from simulation and a completely
new set of bids are calculated given the new (reduced) set of
contracts.  Not doing so would result in \textit{overfilling} some
contracts, which would never be done in practice. Thus, since there
are a total of $6$ contracts, we solve $6$ instances of Problem
\eqref{eqn:dt_problem_cvx} over the course of each simulation run.


When the problem data (i.e., the supply rate curves $W_j$) have been
estimated from the IPinYou dataset, there is almost no difference
between the results obtained for first or second price cases.  That
is, the solution $(x, \gamma)$ for Problem \eqref{eqn:ct_problem} when
$f = f_{2nd}$ is very nearly optimal for the same problem when
$f = f_{1st}$, and vice versa.  This similarity is explained by the
close connection between the two problems established by Proposition
\ref{prop:connection} -- essentially, the main difficulty is in
finding bids which result in a feasible allocation.  Note however,
that all else being equal, second price auctions are naturally much
cheaper than first price auctions; in our case, the total spend in a
first price auction is on average $60\%$ greater than in a second
price auction.  However, this obviously does not account for the
incentive differences of participants in different auction settings.

Synthetic examples where the solutions between first and second price
auctions differ significantly can be constructed and revolve around
the function $g(x, t)$, identified in Example \ref{ex:pseudo_bids},
which is used in the mapping from the constant pseudo bids (dual
multipliers) into (actual) bids $x(t)$.  Since in the second price
case $g^{-1}_{2nd}(p, t) = p$ for every $t$, the most striking
differences occur when $g$ fluctuates rapidly with time.  This
corresponds to the case where the marginal cost of production
$f'_{1st}(x, t) / W'(x, t)$ (if the curves are differentiable in $x$)
vary rapidly, as can be seen in Example \ref{ex:pseudo_bids}.  Major
fluctuations of this sort are not observed in our dataset.  We also
note that examples where the solutions differ, but the supply rate
curves are time homogeneous (i.e., $W_j(x, t) = W_j(x)$) can also be
constructed.

A summary of these numerical results is provided in Table
\ref{tab:results_summary}.  The simulation is carried out where the
\textit{true} auction mechanism is a first price auction.  The first
and second row in the table corresponds to a cost metric (relative
cost per item acquired) with the third row measuring the mean
fulfillment.  The first row measures the average spend per item over
simulations that \textit{actually} fulfilled all contracts (defined as
reaching at least $98\%$ fulfillment) and the second row measures the
average spend per item over all simulations.  The first two columns in
the table correspond to the two possible cost functions used in
solving Problem \eqref{eqn:dt_problem_cvx} (either a first or second
price cost) and the third column provides a p-value comparing the
significance of the differences in the other two columns.  The cost is
measured relative to the first column.

It would be expected that using a cost function which corresponds to
the true auction mechanism would result in improved performance.
However, no real difference is observed.  These results indicate that
in practice, regardless of whether the true auction mechanism is first
price or second price, the bidder can bid \textit{as if} it were a
second price auction.  This is advantageous not only for simplicity,
but because Problem \eqref{eqn:ct_problem_cvx} is \textit{always}
convex for $\Lambda = \Lambda_{2nd}$, but not necessarily for
$\Lambda_{1st}$, see Section \ref{sec:convex_acquisition_costs}.

\begin{table}
  \centering
  \captionsetup{type=table}

  \begin{tabular}{lrrr}
    \toprule
    & \multicolumn{2}{c}{\eqref{eqn:ct_problem} Cost Function} & \\
    \cline{2-3}
    Performance Metric &  $f = f_{1st}$ &  $f = f_{2nd}$ &  t-test $p$ \\
    \midrule
    Spend / Item (fulfilled) &  $1$ &  $0.996$ &  $0.952$ \\
    Spend / Item (total)     &  $1$ &  $1.02$ &  $10^{-6}$ \\
    Mean Fulfillment \%       &  $0.895$ &  $0.895$ &  $0.777$ \\
    \bottomrule
  \end{tabular}

  \caption{Summary of Simulation Results (N = 250)}
  \label{tab:results_summary}

  {Relative average spend and mean normalized contract fulfillment in
    a monte carlo simulation of a first price auction (rows) for
    algorithms based on first price and second price cost functions
    (columns).  (total) indicates averages taken over the entire set
    of simulations, and (fulfilled) indicates averages over fulfilled
    contracts only.  The performance between the two algorithms is
    nearly identical.  The p-values are obtained from paired sample
    t-tests \texttt{scipy.stats.ttest\_rel} and independent sample
    t-tests \texttt{scipy.stats.ttest\_ind} in the case of the first
    row (since there is an unequal number of samples in that case).}
\end{table}

The fulfillment row of Table \ref{tab:results_summary} indicates the
mean (over the $250$ simulations) of the average (over the $6$
contracts) percentage fulfillment.  That is, if we let $c_i^{(k)}(t)$
be the cumulative number of items acquired for contract $i$ by time
$t$ in the $k^{\text{th}}$ simulation, we are reporting the mean over
$k$ of

\begin{equation}
  \label{eqn:avg_fulfillment}
  C^{(k)}_{\text{avg}} \defeq \frac{1}{N} \sum_{i = 1}^N \frac{c^{(k)}_i(T_i)}{C_i}.
\end{equation}

We turn to a further analysis of fulfillment in the next section.

\subsection{Improving Fulfillment}
\label{sec:improving_fulfillment}
Table \ref{tab:results_summary} indicates that, on average, contracts
are only about $90\%$ fulfilled by their deadline.  This is an
important issue, but is not surprising because supply curves are
averages while actual items arrive randomly.  The theory that has been
developed is thus optimal for the average case.  Directly addressing
the stochastic nature of the problem, e.g., via a dynamic stochastic
model, is one possible alternative.  However, such models are
generally intractable, and do not easily lead to practical bidding
algorithms.  On the other hand, while staying within an average
case framework, there are at least two effective methods for improving
upon fulfillment: receding horizon updates, and risk-aware supply
adjustments, which we consider in turn.

\subsubsection{Receding Horizon}
\label{sec:receding_horizon}
Suppose that at time $t$ there has been $c_i(t)$ supply acquired for
contract $i$.  This is useful information which can be incorporated
back into updated supply curve estimates, as well as into the
calculation of future bids and allocations.  Since we are assuming the
use of only crude supply curve estimates from historical data, we
consider only how to incorporate this information into new bid
calculations.  Most straightforwardly, a new instance of Problem
\ref{eqn:ct_problem} can be solved for the updated contracts
$\{(\A_i, C_i - c_i(t), T_i - t)\}_{i = 1}^N$, and this bid
re-calculation carried out at regular time intervals.  This method of
adaptation is referred to as a \textit{receding horizon} method
\cite{kwon2006receding} and accounts for unexpected supply shortfalls
by increasing the bid (which gets us closer to complete fulfillment),
and decreases the bid when unexpected supply surpluses are encountered
(which reduces costs).

\subsubsection{Supply Inflation}
In addition to a receding horizon, it is important to account for
uncertainty in supply curve estimates, as well as the random dynamics
intrinsic to the market.  Our main algorithm does not endogenously
account for the possibility of supply shortfalls, (it pertains only to
an \textit{average} case scenario) but a parameter, $\delta$, can be
appropriately tuned to inflate the supply.

Concretely, if $C$ is the required supply, we can instead \textit{aim}
at an inflated supply target by solving Problem
\eqref{eqn:dt_problem_cvx} with an inflated supply constraint
$(1 + \delta) C$.  The parameter $\delta$ can be tuned via
cross-validation procedures, or calculated to correspond with a model
of estimation and market uncertainty.  This adjustment has the effect
of front-loading the acquisition of supply, a result similar to that
of \cite{grislain2019recurrent}.  In combination with a receding
horizon, excess spending will also be limited.

In order to justify this adjustment, we provide a simple example
model.  Suppose that $M = N = 1$ (i.e., there is a single item type
and a single contract), that we have the contract $(\{1\}, C, T)$, and
that the supply curve is now the \textit{random and time-independent}
quantity $\Wb(x)$ with $\E \Wb(x) = W(x)$, for some known $W(x)$.
Then, given a risk tolerance parameter $\epsilon \in (0, 1)$ we seek a
solution to the following problem

\begin{equation}
  \label{eqn:problem_MN1}
  \begin{aligned}
    \underset{x}{\text{minimize}} \quad & T f(x)\\
    \textrm{subject to} \quad & \P\bigl(T \Wb(x) \ge C \bigr) \ge 1 - \epsilon, \\
  \end{aligned}
\end{equation}

where $f(x)$ is the expected cost per unit time in the auction when
bidding $x$.  Since the objective function is monotone increasing, the
optimal solution to the problem is the smallest $x$ which is feasible
and we have established the following proposition.

\begin{theoremEnd}[normal]{proposition}
  \label{prop:MN1_soln}
  Suppose that $f(x)$ is monotone non-decreasing.  Then, the optimal
  solution of Problem \eqref{eqn:problem_MN1} is given by

  \begin{equation}
    x^\star(\epsilon) = \mathrm{min}\{x\ |\ \P\bigl(\Wb(x) \ge C_i / T \bigr) \ge 1 - \epsilon\}.
  \end{equation}
\end{theoremEnd}

In principle, this result covers every probabilistic model for the
distribution of $\Wb$ including the accounting of uncertainties in
prices, item arrival rates, and estimation inaccuracies.  Moreover,
many such models for the function
$x \mapsto \P\bigl(\Wb(x) \ge C_i / T \bigr)$ can be expected to be
monotone non-decreasing in $x$ (similarly to $f$), in which case the
actual computation of $x^\star(\epsilon)$ poses no difficulty.  It
will be seen in the following example that plausible probabilistic
models can reduce into the framework of supply inflation, i.e., aiming
for some $(1 + \delta) C$.

\begin{theoremEnd}[normal]{example}[Poisson Model]
  Suppose that $\Wb(x) \sim \text{Po}(W(x))$ has a Poisson
  distribution with mean $W(x)$.  Although proposition
  \ref{prop:MN1_soln} still applies to this model, and a computational
  solution is easy to obtain, we will make use of the Poisson Chernoff
  bound to find an illustrative approximation $\tilde{x}(\epsilon)$ in
  closed form.  Replacing the probabilistic constraint in Problem
  \eqref{eqn:problem_MN1} with the right hand side of the bound

  \begin{equation*}
    1 - \P\bigl(\Wb(x) \ge C/T \bigr) \ge 1 - \frac{(e W(x))^{C/T} e^{-W(x)}}{(C/T)^{C/T}},
  \end{equation*}

  which is valid as long as $W(x) \ge C / T$, results in a
  \textit{more} stringent requirement. Thus, a feasible point
  $\tilde{x}(\epsilon)$ for the problem can be found by solving the
  following inequality:

  \begin{align*}
    \frac{(e W(x))^{C/T} e^{-W(x)}}{(C/T)^{C/T}} &\le \epsilon\\
    \overset{(a)}\iff x &\ge W^{-1}\Bigl(-\frac{C}{T}\mathcal{W}_{-1}\bigl(-\frac{1}{e}\epsilon^{T / C}\bigr)\Bigr),
  \end{align*}


  where in $(a)$ $\mathcal{W}_{-1}$ is the lower branch of the
  Lambert-W function, i.e., the inverse function of $x \mapsto xe^x$ ,
  which is decreasing and negative \cite{corless1996lambertw}.  Since
  $-\mathcal{W}_{-1}\bigl(-\frac{1}{e}\epsilon^{T / C}\bigr) \ge 1,$
  this has the form of a \textit{multiplicative} inflation of the
  required supply, and can be written as $(1 + \delta)C$ where
  $\delta = -\mathcal{W}_{-1}\bigl(-\frac{1}{e}\epsilon^{T / C}\bigr)
  - 1$.
\end{theoremEnd}

\begin{figure}
  \centering
  \includegraphics[width=0.75\textwidth]{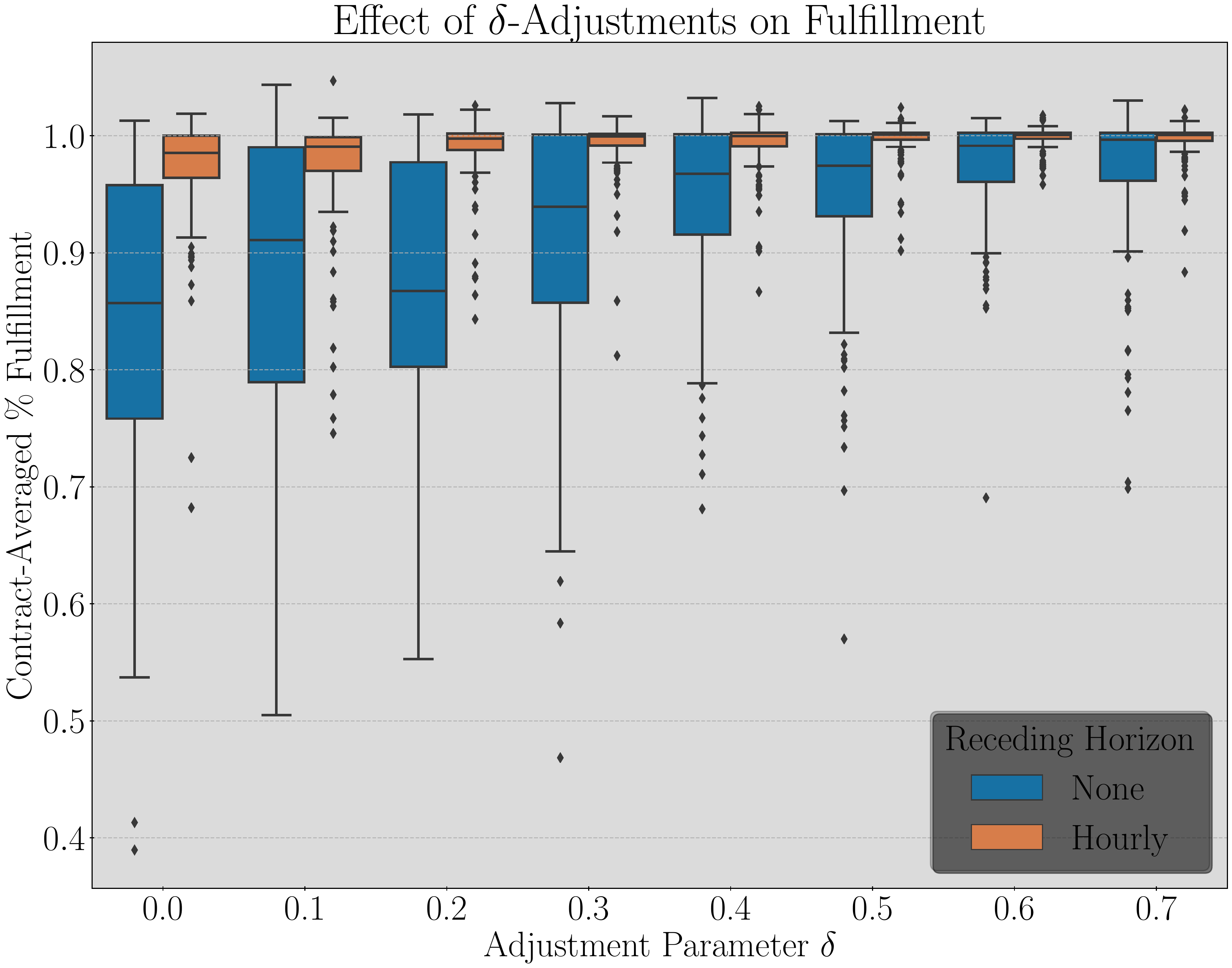}
  \caption{Risk Adjustment Simulations (Fulfillment)}
  \label{fig:simulation_risk_adjustment}
\end{figure}

\begin{figure}
  \centering
  \includegraphics[width=0.75\textwidth]{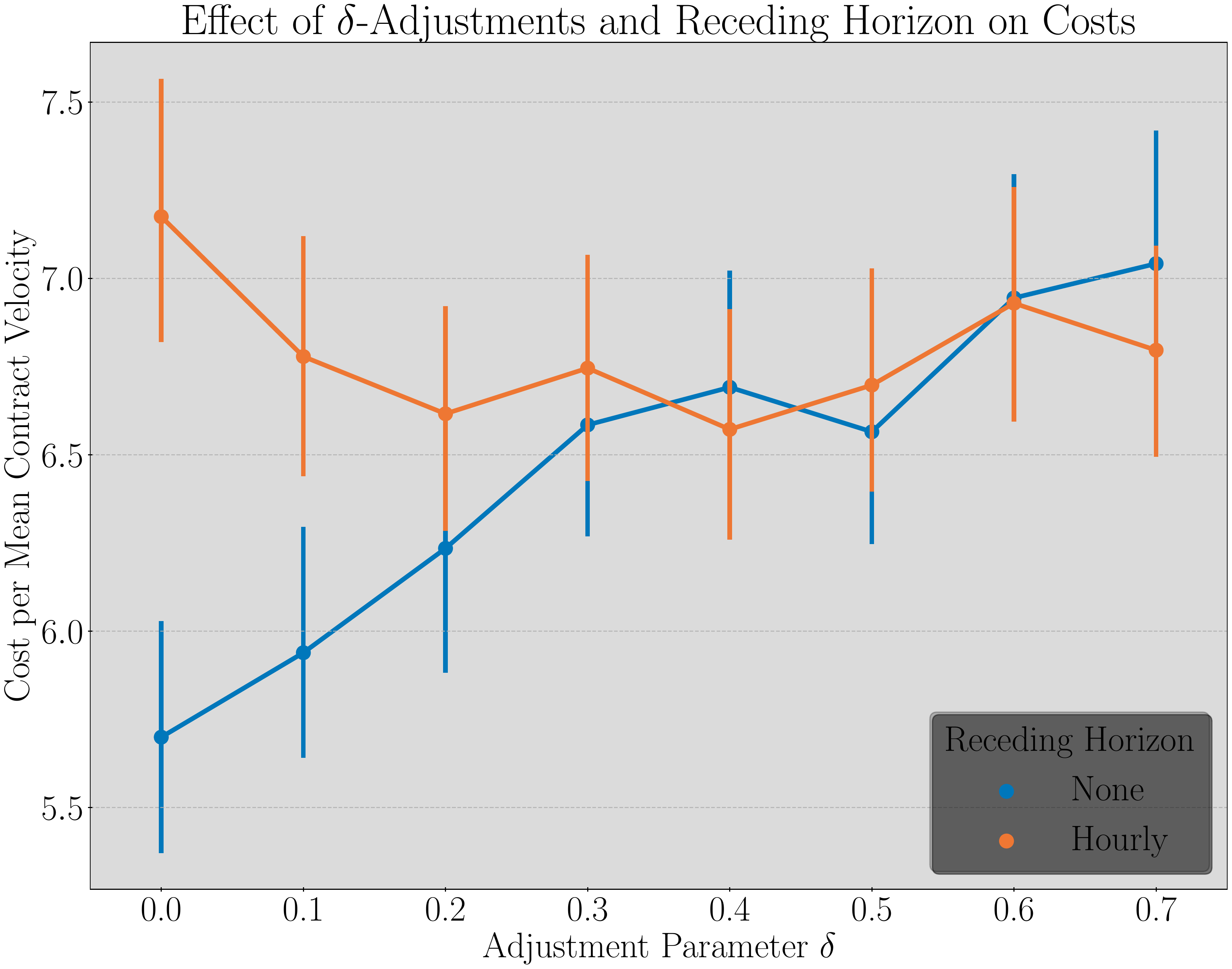}
  \caption{Risk Adjustment Simulations (Fulfillment)}
  \label{fig:simulation_risk_adjustment_cost}
\end{figure}

\subsubsection{Sensitivity Simulations}
In order to illustrate the effects of a $\delta$ risk adjustment and
the receding horizon re-calculations, we have carried out simulations
similar to those of Section \ref{sec:simulation}, however with smaller
contracts $N = M = 3$.  The results are summarized by Figures
\ref{fig:simulation_risk_adjustment} and
\ref{fig:simulation_risk_adjustment_cost}.

Figure \ref{fig:simulation_risk_adjustment} provides a box plot
comparing the parameter $\delta$ (applied to inflate the supply of
each contract by the same proportion) to the average contract
fulfillment (see $C_a^{(k)}$, Equation \eqref{eqn:avg_fulfillment}).
As is expected, the proportion of contracts which are completely
fulfilled increases with $\delta$.  In addition, the plot provides
pairs of boxes comparing the results when the bids are updated only
after a contract is fulfilled (this is the minimal update schedule
that doesn't overfill contracts), and when the bids are also updated
every hour of simulation time (typically, about 20 updates are
calculated over the simulation).  An hourly receding horizon has a
dramatic affect on the fulfillment -- even without a risk adjustment,
more than half of the contracts are fulfilled to at least $98\%$, and
with a risk adjustment, almost all contracts are completely fulfilled.

Figure \ref{fig:simulation_risk_adjustment_cost} provides a line plot
similar to Figure \ref{fig:simulation_risk_adjustment} except the
ordinate now quantifies the cost paid for items.  Points are mean
values across $500$ simulations and error bars are $98\%$ confidence
intervals.  In order to create a reasonable cost metric which is
comparable across different contracts, we have normalized the total
cost of attempting to fulfill a contract by that contract's ``mean
velocity'' requirement.  To understand what we refer to as the
velocity requirement, suppose that we are obligated to obtain $C$
items in $T$ time.  Then, we define the velocity of this requirement
by $v = C / T$.  We take the mean velocity of the whole contract to be
the average of the velocities of each individual requirement in the
contract (in this case, over the $N = 3$ requirements).  To merely
normalize by, e.g., the total number of items, would not result in a
fair comparison since one contract may require obtaining the same
number of items in less time (since they are generated randomly),
which necessitates paying higher prices.

The main conclusions to be drawn from
\ref{fig:simulation_risk_adjustment_cost} are as follows.  Firstly,
without a receding horizon, there is naturally a tendency for costs to
increase, since the supply acquisition is front-loaded.  Secondly, as
$\delta$ increases, the same upward pressure on costs is applicable as
in the case without any receding horizon, but at the same time, for
contracts which are difficult to fulfill, the receding horizon will
drastically increase bids, which also increases the cost.  However,
for contracts which are easy to fulfill, or which get close to
fulfillment early in the bidding period, the receding horizon smooths
out the acquisition rate by reducing the bid, and therefore reducing
costs.  These are competing effects, but the use of the receding
horizon is still able to reduce costs when $\delta$ is large; an
ultimately small cost for achieving near complete fulfillment.

In summary, the receding horizon is able to achieve a much higher
fulfillment proportion by increasing the bids (and incurring higher
costs) for contracts which are difficult to fulfill (e.g., when supply
rates are overestimated), while at the same time reducing the costs of
fulfilling the remaining contracts by smoothing out bids that are too
large.

\section{Conclusions}
\label{sec:conclusion}
This paper has studied the problem of minimum cost contract
fulfillment in RTB auction markets.  We have analyzed generalizations
of the model proposed in \cite{marbach_bidding_2020} by considering
both the first and second price auction mechanisms, and by allowing
for general time-varying supply rate curves which can account for
daily and weekly periodic cycles in market data.  Moreover, we have
uncovered the intrinsic convexity of the problem and provided a
natural condition under which a strong duality result holds.  Using
duality, it has been shown that there is a close connection between
the optimal bids $x(t)$ and certain dual variables which facilitates
the complete representation of the optimal function $x(t)$ through the
\textit{finite} vector $\rho \in \R^N_+$ of dual variables.  This
close connection motivates the terminology ``pseudo-bids'' for the
dual variables $\rho$.  In the first price case, the optimal bids are
obtained from $\rho$ by mapping them through a time-varying function
quantifying the marginal costs of acquiring additional items; in the
second price case (but \textit{not} the first), the optimal bids are
given exactly by $\rho$, which implies the optimal bids in this case
are piecewise constant functions of time.

Further consequences of duality have lead to the conclusion (backed up
with experimental simulation evidence) that there is in fact very
little difference between the optimal solution in the first price or
second price cases.  This result has the important practical
implication that, regardless of the type of auction one is
participating in, they may as well \textit{assume} that they are
participating in a second price auction, which has been seen to
\textit{always} lead to a convex problem.  Indeed, in second price
auctions, the optimal bids are \textit{constant} (corresponding
exactly to the dual pseudo-bids $\rho$) and therefore the original
time-independent formulation of \cite{marbach_bidding_2020} largely
covers the general time-dependent case after calculating suitable time
averages.

Finally, Section \ref{sec:computational_methods} develops methods for
implementation and reports the results of simulation studies.  A
discretization method for Problem \eqref{eqn:ct_problem_cvx} is
studied in Section \ref{sec:approximation} and we obtain a convergence
rate for this discretization.  The discretized problem is used in an
empirical Monte Carlo study in Section \ref{sec:simulation}, which
demonstrates the similarity between optimal bidding in first and
second price auctions.  Further simulations in Section
\ref{sec:improving_fulfillment} demonstrate the effectiveness of
$\delta$ supply inflation and receding horizon bid recalculation for
ensuring that contracts are actually fulfilled in practice.
Additional information regarding the practical implementation of these
methods is provided in the appendix, with the basic ideas being
summarized in Algorithm \ref{alg:eli5_bidding}.

\printbibliography
\clearpage

\appendix
\section{Supply Rate Curve Estimation}
\label{sec:supply_curve_estimation}
In practice, the functions $W(x)$ will be estimated from available
historical data.  Therefore, depending on the method used to carry out
this estimation, the resulting $\Lambda(x)$ is not necessarily
guaranteed to be convex in the first price case (recall Proposition
\ref{prop:convex_acquisition_costs} requires $2$-concavity of $W$).
If the estimate of $W(x)$ is simply carried out by fitting a
particular parameterized distribution (e.g., a normal approximation)
to a dataset, then there is unlikely to be any issue since most
distributions commonly employed for this purpose do in fact have
log-concave cumulative distribution functions.  Moreover, since $W(x)$
arises through an auction process, extreme value theory and the
Fisher-Tippett-Gnedenko Theorem may motivate the belief that $W(x)$
should tend to be close to a Weibull distribution, which is
log-concave at least for some parameter ranges.

However, our empirical data (see also \cite{liao2014ipinyou,
  zhang2014real}) suggests that such simple models are \textit{not}
good estimates for supply rate curves, and that the curves have a tendency
towards some multi-modality.  For this reason, the methods of Section
\ref{sec:simulation} make use of Kernel density estimates (KDE).  A
Gaussian KDE model for $W(x)$ is natural because it ensures
smoothness, that $W > 0$, and the bandwidth can be chosen to
correspond to the level of bid noise (or the level of bid noise can be
chosen from the optimal KDE bandwidth).

Unfortunately, KDE estimates of $W(x)$ from data need not be (and
often aren't) log-concave or $\alpha$-concave.  In order to deal with
this problem, we consider calculating a convex and piecewise affine
majorant of the function $\Lambda(s)$, which will transform Problem
\eqref{eqn:dt_problem_cvx} into a linear program.  A similar
minorizing envelope can also be calculated through the methods of
\cite{oberman2007convex}, and a piecewise minorant thereof computed by
outer linear approximations.  There is no obvious reason to prefer one
approximation over the other and our experience has not demonstrated a
clear benefit either way.

Alternatively, taking the log-concave envelope of $W(x)$ through
similar methods is guaranteed to result in convex acquisition
functions.  The considerations of the previous paragraph suggest that
we should at least expect the KDE estimates of $W$ to be ``almost''
log-concave, and indeed, this is what we have observed in our own
experiments; the convex envelopes are only slight perturbations of the
original supply rate curve estimate.

\subsection{Piecewise Affine Approximation}
\label{sec:pw_aff_majorant}
Let us denote by $\Lt(x)$ the acquisition cost function attained from
an estimated supply rate curve and by $\Lambda^{\mathsf{U}}(x)$ the minimal
convex majorant $\Lt(x)$, i.e.,

\begin{equation}
  \label{eqn:log_concave_minorant}
  \Lambda^{\mathsf{U}}(x) = \text{inf}\{\lambda(x)\ |\ \lambda(x) \ge \Lt(x), \lambda \text{ is convex and monotone increasing}\}.
\end{equation}

We emphasize that $\lambda$ must be monotone increasing, but this would also
follow as a consequence of the monotonicity of $\Lt$.  The maximal
minorant can be defined similarly.  Moreover, $\alpha$-concave
envelopes can be calculated by requiring that $\ell_\alpha \circ \lambda$ is
convex.

A piecewise affine approximation of $\Lambda^{\mathsf{U}}$ can be
found by discretizing a compact interval $[a, b] \subset \R$ into
$n + 1$ points $x_0, x_1, \ldots, x_n$ and solving the following
convex quadratic program where convexity and monotonicity are enforced
via finite differences

\begin{equation}
  \label{eqn:lower_envelope}
  \begin{aligned}
    \underset{\lambda}{\mathrm{minimize}} \quad & \frac{1}{n + 1}\sum_{i = 0}^n\bigl(\lambda_i - \Lt(x_i))\bigr)^2\\
    \mathrm{subject\; to} \quad & \lambda_i \ge \Lt(x_i)\\
    \quad & \lambda_{i} - \lambda_{i - 1} \ge 0\ \forall i \in [n]\\
    \quad & \lambda_{i + 1} - 2\lambda_i + \lambda_{i - 1} \ge 0\ \forall i \in [n - 1].
  \end{aligned}
  \tag{$E^{\mathsf{U}}$}
\end{equation}

An accurate approximation of the convex majorant is recovered via
linearly interpolating $\lambda_i$.  In fact, $\lambda_i$ will result in a
\textit{strictly} monotone function (and therefore a continuous
inverse) whenever $\Lt$ is strictly monotone.

It is important that the developments in Section
\ref{sec:optimal_bidding_in_ct} do not make any assumptions regarding
the differentiability of $W(x)$ or $\Lambda(s)$, since if these curves
are corrected through solution of Problem \eqref{eqn:lower_envelope}
to ensure convexity, it is by definition not differentiable.
Moreover, simply interpolating the resulting $\lambda_i$ (e.g. with a
cubic spline) may not be acceptable as, to our knowledge, it is not
possible for this process to maintain simultaneously the monotonicity,
convexity, and minorization of $\Wt$.

\begin{remark}[Sparse Approximations]
  \label{rem:sparse_approx}
  It is desirable to use a fine discretization in Problem
  \eqref{eqn:lower_envelope}, otherwise the resulting function may
  fail to majorize $\Lt$ in regions of high curvature.  However, each
  piecewise segment translates to an additional constraint when
  $\Lambda^{\mathsf{U}}$ is substituted into Problem
  \eqref{eqn:dt_problem_cvx} (see Section
  \ref{sec:linear_approximation}), which may become overly burdensome
  for large problems.  Therefore, it may be desirable to use a coarser
  approximation obtained by linearly interpolating samples of
  $\Lambda^{\mathsf{U}}(s)$.  Since $\Lambda^{\mathsf{U}}(s)$ is
  convex, this process is guaranteed to produce another piecewise
  affine convex function which further majorizes
  $\Lambda^{\mathsf{U}}(s)$.
\end{remark}

\subsection{An Example}
We consider an illustrative example of calculating log-concave
envelopes of the supply rate curve in a simple market model.  We suppose
that each participant is characterized by a bid and rate pair
$(b_i, r_i)$ indicating that they will bid $b_i$ with probability
$r_i$ on any arriving item.  We sample $\{(b_i, r_i)\}_{i = 1}^{30}$
randomly as $b \sim \text{exp}(0.5)$ and $r_i \sim \beta(11.1, 10)$
which represents a market with $30$ participants whose average bid is
$0.5$ and have an average probability of
$(11.1 - 10) / 11.1 \approx 0.1$ of bidding.

The \textit{bid landscape} in this situation is given by

\begin{equation*}
  W(x) = \prod_{i: b_i > x}(1 - r_i),
\end{equation*}

indicating the probability of winning an item if the bid $x$ is
placed.

\begin{figure}
  \centering
  \includegraphics[width=0.75\linewidth]{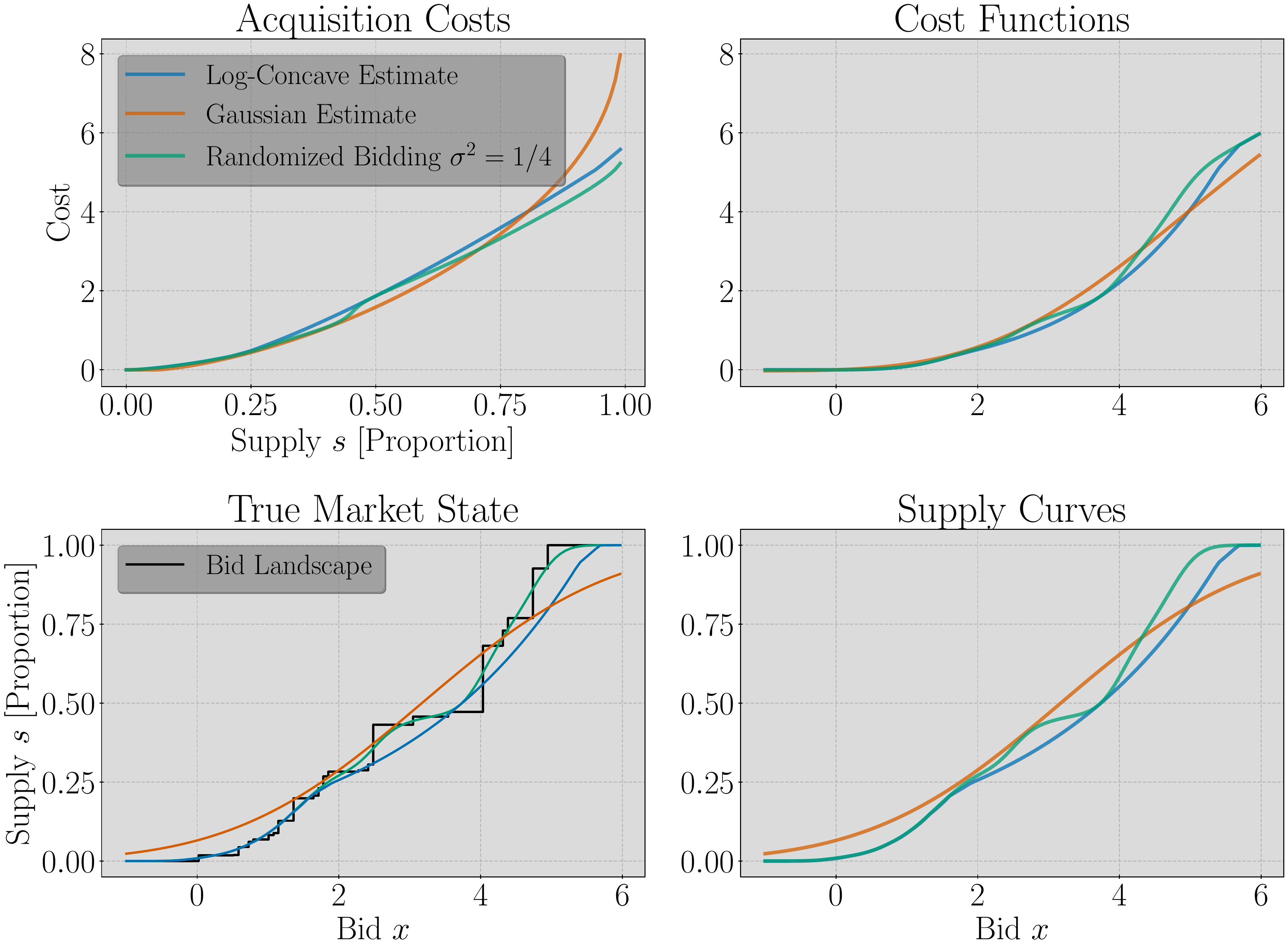}
  \caption{Example Supply Rate Functions}
  \label{fig:example_functions}
  {Comparison of different methods of estimating supply rate curves.
    Lower left: Comparison of KDE smoothing, the maximal log-concave
    minorant thereof, and a Gaussian c.d.f. (fit by moment matching)
    overlaid upon a true market state.  Lower right: The three
    corresponding supply rate curves.  Upper right: Corresponding cost
    curves.  Upper left: Corresponding acquisition cost functions
    where we see that KDE smoothing does not lead to convexity, and
    that a Gaussian estimate is not a consistent minorant or
    majorant.}
\end{figure}

We let $\Wt(x)$ be a KDE smoothed (with $\sigma^2 = 1/4$) version of
$W$ which corresponds either to the true supply rate curve under randomized
bidding, or a reasonable estimate (from historical data) thereof.  We
denote supply rate curve estimates $\Wt^{\mathsf{L}}(x)$ and
$\Wt_\mathcal{N}$ which are derived from $W(x)$ by solving Problem
\eqref{eqn:log_concave_minorant} for the function
$-\text{log} \circ W$ and moment matching a Gaussian c.d.f.,
respectively.  Note that this procedure produces \textit{minorants} of
the supply rate curve (and therefore majorants of the acquisition cost
curve), since using a \textit{convex} majorant procedure results in
minorants of \textit{concave} functions.  Figure
\ref{fig:example_functions} plots examples of these functions and
their associated cost and acquisition counterparts.

The examples of Figure \ref{fig:example_functions} are chosen to
deliberately exaggerate the differences between the supply curve
estimates and the envelopes.  For the simulation examples of Section
\ref{sec:simulation} the two curves are often indistinguishable.

\subsection{Linear Approximations of Primal Problem}
\label{sec:linear_approximation}
From Section \ref{sec:approximation} we have the time discretized
primal problem \eqref{eqn:dt_problem_cvx}:

\begin{equation}
  \begin{aligned}
    \underset{s, r}{\mathrm{minimize}} \quad & \sum_{j = 1}^M \sum_{k: \Tb_k < T^j} \Lb_{jk}(s_j[k])\\
    \mathrm{subject\; to} \quad & \sum_{j \in \A_i} \sum_{k: \Tb_k < T^j} \Delta_k r_{ij}[k] \ge C_i\\
    \quad & \sum_{i \in \B_j \cap \Tk} r_{ij}[k] = s_j[k]\\
    \quad & r_{ij}[k] \ge 0.
  \end{aligned}
  \tag{$\PK{K}$}
\end{equation}

Suppose that the functions $\Lambda_j(s, \widetilde{T}_k)$ have
piecewise affine approximations
$\widehat{\Lambda}_k(s) = \underset{h \in H_{jk}}{\text{max}}
\bigl(m_{hj}[k]s + b_{hj}[k] \bigr)$.  Then, by introducing additional
variables $\alpha_{jk}$ we can reformulate $\PK{K}$ in epigraph form
as a linear program

\begin{equation}
  \begin{aligned}
    \underset{s, r}{\mathrm{minimize}} \quad & \frac{1}{2}\sum_{j = 1}^M \sum_{k: \Tb_k \le T^j} \Delta_k (\alpha_{jk} + \alpha_{j, k - 1})\\
    \mathrm{subject\; to} \quad & \sum_{j \in \A_i} \sum_{k: \Tb_k \le T^j} \Delta_k r_{ij}[k] \ge C_i\\
    \quad & m_{hj}[k]s_j[k] + b_{hj}[k] \le \alpha_{jk}\\
    \quad & \sum_{i \in \B_j \cap \Tk} r_{ij}[k] = s_j[k]\\
    \quad & r_{ij}[k] \ge 0,
  \end{aligned}
\end{equation}

which is the formulation we have employed for our simulations in
Section \ref{sec:simulation}.

\section{Simulation (Additional Details)}
In this section we provide additional details on the methods used to
produce the results of Section \ref{sec:simulation}.

\subsection{Estimating Supply Rate Curves}
\label{sec:estimating_supply_curves}
\begin{figure}
  \centering

  \begin{subfigure}[b]{0.4\textwidth}
    \includegraphics[width=\textwidth]{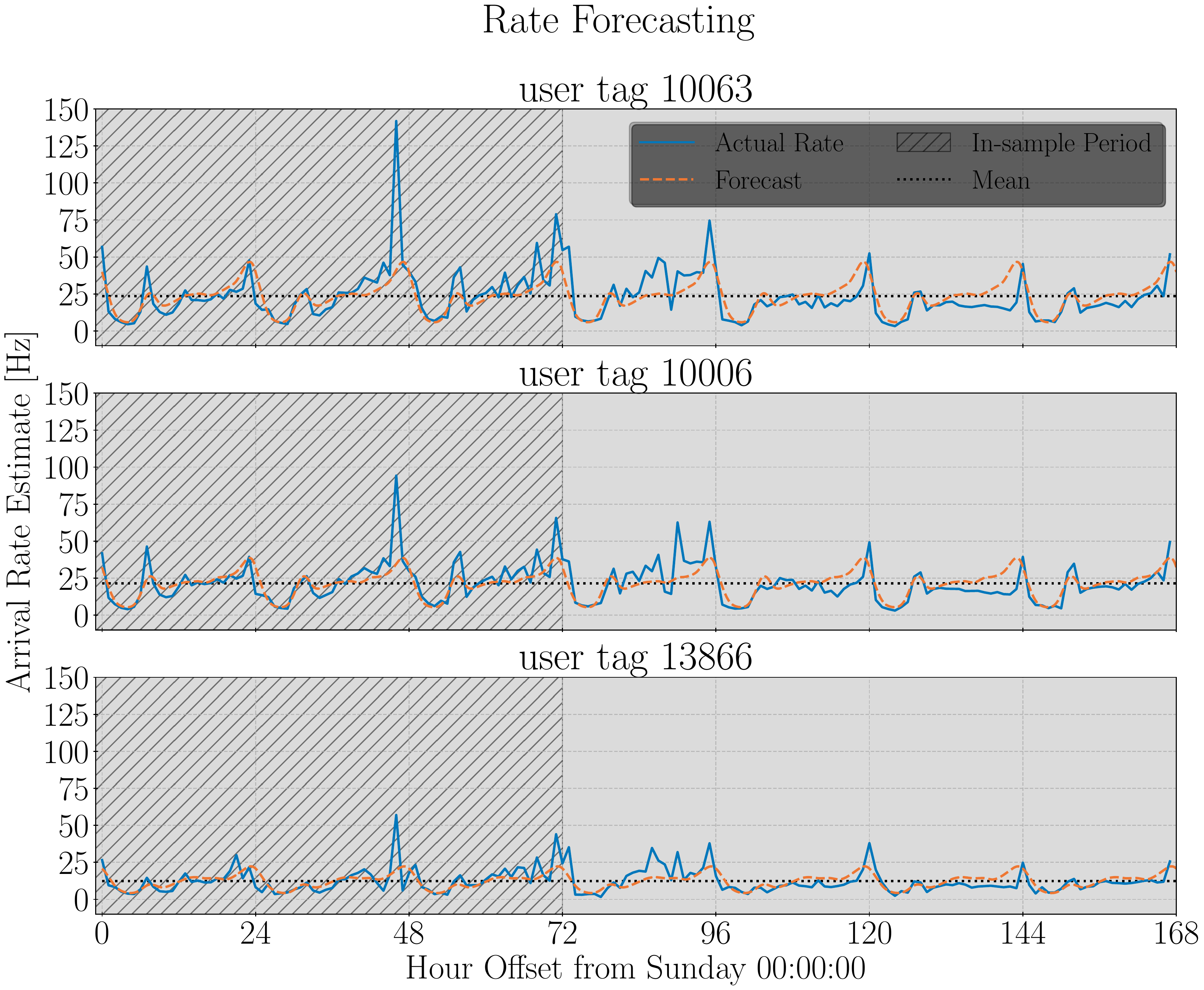}
    \caption{Supply Rate $\lambda(t)$}
    \label{fig:supply_rate_estimate_apx}
  \end{subfigure}
  \begin{subfigure}[b]{0.4\textwidth}
    \includegraphics[width=\textwidth]{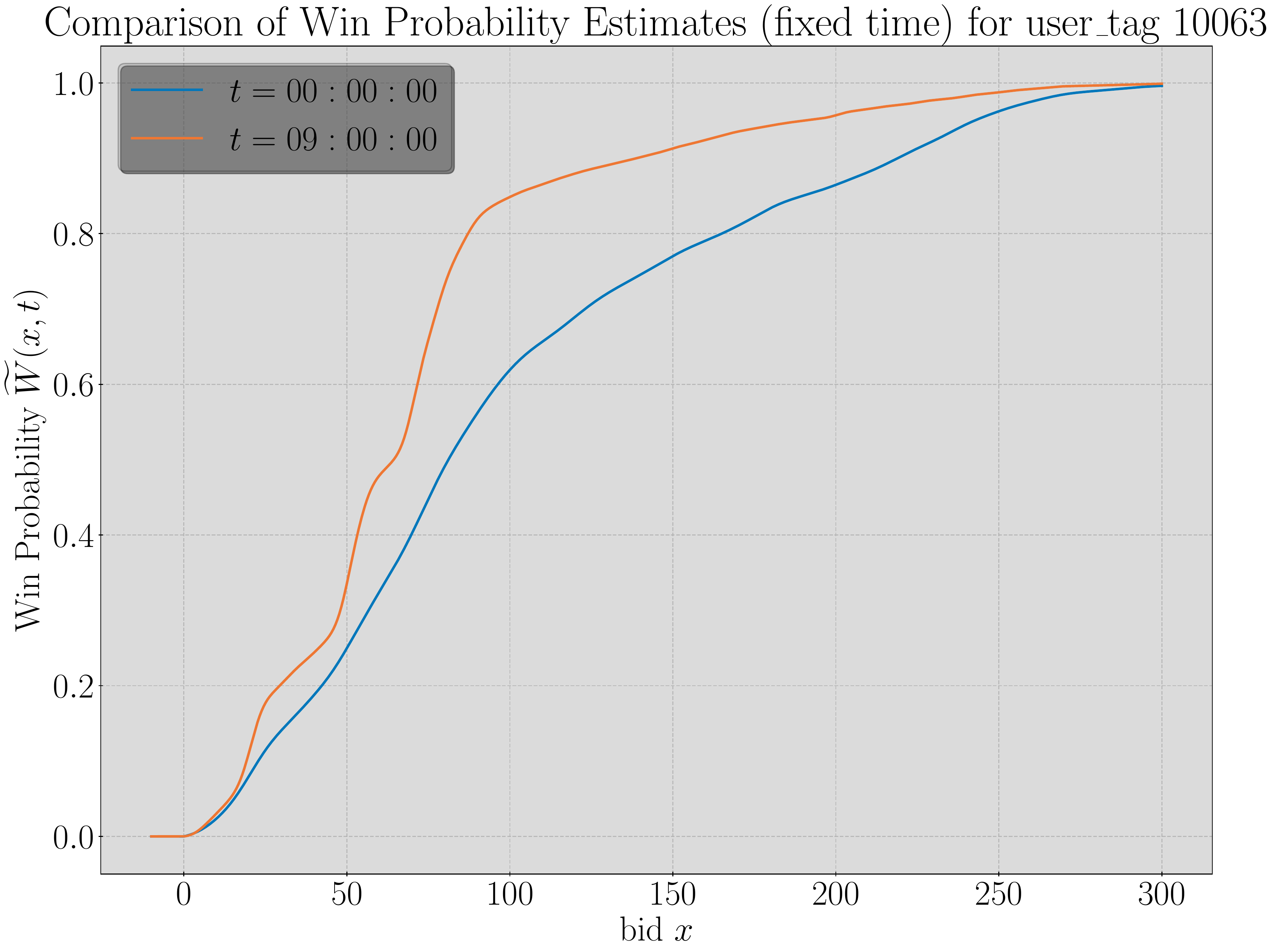}
    \caption{Win Probability Estimates}
    \label{fig:win_prob_estimate_apx}
  \end{subfigure}
  \caption{Estimated Supply Rate Curves and Costs}
  \label{fig:supply_curve_estimates_apx}


  {Cost and supply rate curves estimated from iPinYou data from the
    first 3 days of season 2.  $(a)$ Item arrival rates and the
    corresponding forecasts.  The hatched region indicates an
    in-sample period with the remainder being out-of-sample.  Our
    simulations run on similar 3 day periods with a 12 hour sliding
    window for a total of 9 periods of 72 hours each.  $(b)$ Estimates
    of the win probability function $\widetilde{W}(x, t)$ for
    $t = 00:00:00$ (blue) and $t = 09:00:00$ (red).  We compare
    Gaussian KDE (solid line) with a parametric Exponential CDF
    (dashed line).}
\end{figure}

The iPinYou dataset consists of impression data derived from a real
DSP and includes information about bidding prices, market prices, and
user characteristics.  We focus on the \textit{season two} data (a
week long period $2013$-$06$-$06$ to $2013$-$06$-$12$).  In all cases,
our supply rate curve estimates are $24$h-periodic in time and
therefore account for the natural daily (but not weekly) cycles in
prices and arrival rates.  Since this paper does not focus on the
estimation of supply rate curves, we apply a simple estimation
procedure using the entire dataset as input.  Though this has the
effect of leaking some information from the future, the estimation of
supply rate curves is not subject to optimization, limiting the impact
of this leakage.  Moreover, the dataset is averaged into a single
$24$h periodic function and extended through periodicity.  It is
reasonable to believe that the previous week's (out of sample) data
would provide similar results.  Figure
\ref{fig:supply_curve_estimates_apx} provides an illustration of estimated
supply rate curves where only $72$h of data is used to forecast the
remaining $96$h for purposes of illustrating a case where historical
data is used in this manner to forecast future trends.

Provided with the iPinYou dataset is a \texttt{user\_tag} which,
according to \cite{zhang2014real} is ``[a segment] in iPinYou’s
proprietary audience database''.  We therefore use the
\texttt{user\_tag} property as the ``item types'', focusing on the
five most common tags: \texttt{10063}, \texttt{10006},
\texttt{13866}, \texttt{10024}, and \texttt{10083}.

\subsection{Estimating the Supply Rate}
In order to estimate the supply rate $\lambda_j(t)$ for each user tag,
we have taken the inverse of the average of the time differences
between arrival instants in each hour of the day, after removing
outliers.  Calculating the number of arrivals over an hour long period
is not adequate as there appear to be large consistent gaps in arrival
times: we suspect that the dataset was subsampled prior to being
released.

This calculation results in estimates
$\lambda_j[0], ..., \lambda_j[23]$, with time denoted in hours.  The
continuous estimate was subsequently formed by smoothly interpolating
between these points with a $24-$periodic boundary, resulting in a
function $\tilde{\lambda}_j(t)$ defined on $[0, 24]$.  A forecast for
the average supply rate at time $t$ is obtained via
$\lambda_j(t) = \tilde{\lambda}(t\; \text{mod}\; 24)$.

An illustrative example for tags \texttt{10063}, \texttt{10006}, and
\texttt{13866} is provided in Figure \ref{fig:supply_rate_estimate_apx}.

\subsection{Estimating Win Probabilities}
Similarly to the supply rate estimates, we estimate an average win
probability function for each $t \in {0, \ldots, 23}$ and then
smoothly interpolate along $t$ to estimate a $24-$periodic function
$\widetilde{W}_j(x, t)$ indicating the probability of winning an
impression of type (user tag) $j$ arriving at time $t$ given a bid
$x$.

The estimate of $x \mapsto \widetilde{W}_j(x, t)$ is obtained by
smoothing the histogram with a Gaussian kernel (bandwidth chosen
simply by the Normal Reference Rule
\cite[Chap.~6.3]{wasserman2006all}) for each \texttt{market\_price}
data point falling into the hour long window.  The results of this
procedure, as well as a comparison to a parametric estimate with an
Exponential density are given in Figure \ref{fig:win_prob_estimate_apx}.

The \texttt{market\_price} attribute in the dataset corresponds to the
price actually paid in the second price auction.  We have not
accounted for the affects of censoring; since the DSP collected the
dataset with large bids intended to win most impressions that were bid
on, this isn't a significant factor.

\subsection{Cost and Supply Rate Curves}
The supply rate curve $W_j(x, t)$ is simply the product of the supply
rate $\lambda_j(t)$ and the win probability $\widetilde{W}_j(x, t)$.
The cost curves $f^{1st}$ and $f^{2nd}$ is derived from the supply
rate curve using numerical integration and interpolation.

\subsection{Simulating the Bidding Process}
\label{sec:bidding_process_simulation}

The simulations of Section \ref{sec:simulation} are obtained by
storing the hour-by-hour inter-arrival and price data for each item
type $j \in [M]$ and sampling uniformly from these datasets.  At
simulation time $t \in \R_+$ we sample an inter-arrival time
$\Delta t$ and price $P$ from the data for hour
$\lfloor t \rfloor + 1$ with probability $t - \lfloor t \rfloor$ and
otherwise from the data for hour $\lfloor t \rfloor$.  A bid is
solicited from a bidder (an implementation of
\eqref{eqn:dt_problem_cvx}) and if the bid exceeds $P$ the bidder
allocates that item to the fulfillment of a contract.  The simulation
time is them updated to $t + \Delta t$ and the process continues.

\begin{algorithm}
  \SetKwInOut{Input}{input}
  \SetKwInOut{Output}{output}
  \SetKwInOut{Initialize}{initialize}
  \DontPrintSemicolon

  \SetKwFunction{FSample}{Sample-Dataset}

  \BlankLine
  \caption{Bidding Simulation}
  \label{alg:bidding_process}
  
  \Input{A $\text{Bidder}$ derived from Section
    \ref{sec:approximation}.}

  \Output{Recording of Bidder's item
    allocations to process into normalized acquisition curves.}

  \texttt{// Initialize:}\;
  $Q \leftarrow \text{Priority-Queue}([\ ])$ \texttt{  // Sort by time}\;
  $t \leftarrow 0$ \texttt{  // The ``current'' time}\;
  \For{$j \in [M]$} {
    \texttt{  // Sample an inter-arrival time and a price}\;
    $(\Delta t, P) \leftarrow\ $\texttt{Sample-Dataset($t$,$j$)}\;
    $Q$\texttt{.push($(t + \Delta t, P, j)$)}\;
  }

  \BlankLine
  \texttt{// Simulate bidding process:}\;
  \While{$t < T_{\text{end}}$} {
    $t, P, j \leftarrow Q$\texttt{.pop()}\;
    $b \leftarrow \text{Bidder}$\texttt{.solicit\_bid($t$,$j$)} \texttt{  // Ask for a bid on type $j$ at time $t$}\;
    \If{$b \ge P$} {
      $\text{Bidder}$\texttt{.award\_item($t$,$j$)} \texttt{  // Allocate items for winning bids}\;
    }
    $(\Delta t, P) \leftarrow\ $\texttt{Sample-Dataset($t$,$j$)} \texttt{  // Append next $(t, P)$ pair for $j$ to $Q$}\;
      $Q$\texttt{.push($(t + \Delta t, P, j)$)}\;
  }

  \BlankLine
  \SetKwProg{Fn}{Function}{\texttt{($t$,$j$):}}{\KwRet}
  \Fn{\FSample}{
    $p \leftarrow t - \lfloor t \rfloor$\;
    $U \sim \mathcal{U}(0, 1)$  \texttt{  // Interpolate between hours}\;
    \If{$p \le U$}{
      $h \leftarrow \lfloor t \rfloor$\;
    }
    \Else{
      $ h \leftarrow \lfloor t \rfloor + 1$\;
    }
    $\Delta t \leftarrow $\texttt{ Sample-Inter-arrivals(hour=$h$,type=$j$)}\;
    $P \leftarrow $\texttt{ Sample-Prices(hour=$h$,type=$j$)}\;
    \Return $(\Delta t, P)$\;

  }
\end{algorithm} 

\section{Proofs}
\label{sec:proofs}

\subsection{Proofs of Main Results}
\printProofs[main]

\subsection{Additional Proofs}
\printProofs[aux]
\printProofs[aux2]

\end{document}